\documentclass[aps,twocolumn,showpacs,preprintnumbers,amsmath,amssymb]{revtex4-1} 

\usepackage[english]{babel} 

\usepackage{dcolumn}
\usepackage{amssymb}
\usepackage{amsmath}
\usepackage{graphicx}
\usepackage{color}
\usepackage{comment}
\usepackage{lineno}
\usepackage{url}
\usepackage{siunitx}
\usepackage{xcolor} 
\usepackage[linkcolor=black,citecolor=black,urlcolor=black,colorlinks=true]{hyperref}

\newcommand{\wclu}{w_\text{clu}}
\newcommand{\lclu}{l_\text{clu}}
\newcommand{\wnuc}{w_\text{nuc}}
\newcommand{\lnuc}{l_\text{nuc}}

\newcommand{\Nleft}{N_\text{left}}
\newcommand{\Nright}{N_\text{right}}

\newcommand{\xnucz}{x_\text{nuc}^0}
\newcommand{\Dnuc}{D_\text{nuc}}

\begin{document}

\title{Can a flux-based mechanism explain positioning of protein clusters in a three-dimensional cell geometry?}

\author{Matthias Kober}
\email{These two authors contributed equally}
\author{Silke Bergeler}
\email{These two authors contributed equally}
\author{Erwin Frey}
	\thanks{Correspondence: frey@lmu.de}
\affiliation{
Arnold Sommerfeld Center for Theoretical Physics and Center for NanoScience, Department of Physics, Ludwig-Maximilians-Universit{\"a}t M{\"u}nchen, Theresienstra{\ss}e 37, D-80333 Munich, Germany\\
}

\begin{abstract} 
The plane of bacterial cell division must be precisely positioned.
In the bacterium \textit{Myxococcus xanthus}, the proteins PomX and PomY form a large cluster, which is tethered to the nucleoid by the ATPase PomZ and moves in a stochastic, but biased manner towards midcell, where it initiates cell division. 
Previously, a positioning mechanism based on the fluxes of PomZ on the nucleoid was proposed. 
However, the cluster dynamics was analyzed in a reduced, one-dimensional geometry. 
Here we introduce a mathematical model that accounts for the three-dimensional shape of the nucleoid, such that nucleoid-bound PomZ dimers can diffuse past the cluster without interacting with it.
Using stochastic simulations, we find that the cluster still moves to and localizes at midcell. 
Redistribution of PomZ by diffusion in the cytosol is essential for this cluster dynamics. 
Our mechanism also positions two clusters equidistantly on the nucleoid. 
We conclude that a flux-based mechanism allows for cluster positioning in a biologically realistic three-dimensional cell geometry.

\begin{description}
\item[Keywords] 
midcell localization, \textit{Myxococcus xanthus}, cell division, stochastic simulation, \\
flux-based mechanism
\end{description}
\end{abstract}

\maketitle

\section{Introduction}
In bacteria, intracellular positioning of proteins is important for vital biological processes, including localization of the cell division machinery to midcell, as well as chromosome and plasmid segregation. 
The positioning systems responsible often involve P-loop ATPases such as ParA and MinD 
\cite{Gerdes2010, Lutkenhaus2012}. 
These ATPases switch between an ATP- and ADP-bound state, which alters their subcellular localization: the ATP-bound form typically binds as a dimer to the nucleoid or membrane, while the ADP-bound form diffuses freely in the cytosol. 
Activating proteins stimulate the ATPase activity of the ParA-like proteins, which results in detachment of the protein from its respective scaffold in the ADP-bound form. 
Intracellular patterns of these ParA-like ATPases depend on the binding properties of the ADP- and ATP-bound forms, the localization of the stimulating proteins, and the cell geometry \cite{Thalmeier2016}.

Recently, a protein system that includes a ParA-like ATPase has been identified, which regulates the localization of the cell division site in the bacterium \textit{Myxococcus xanthus} \cite{Schumacher2017a, Treuner-Lange2013, Schumacher2017}.
The Pom system consists of three proteins: PomX, PomY and PomZ. 
The ATP-bound ParA-like ATPase PomZ binds as a dimer non-specifically to DNA. 
PomX and PomY form a macromolecular cluster, the PomXY cluster, which is tethered to the nucleoid via PomZ dimers. 
This tethering is transient, since PomX, PomY and DNA synergistically stimulate the ATPase activity of PomZ, which results in two ADP-bound PomZ monomers being released into the cytosol. 
Fluorescence labeling shows that, shortly after cell division, the PomXY cluster moves from a position close to one nucleoid end towards mid-nucleoid in a biased random walk-like movement that depends on PomZ. 
Once at mid-nucleoid, the Pom cluster positively regulates FtsZ ring formation \cite{Schumacher2017a}.

Though the Pom system in \textit{M. xanthus} and the Min system in \textit{Escherichia coli} serve the same function (localization of the cell division plane), they differ in important respects: 
Dimeric MinD-ATP binds to the cell membrane, and MinC inhibits FtsZ ring formation outside of the midcell region, leading to negative regulation of mid-plane localization.  
In this regard, the Pom system is more similar to the Par system for low-copy-number plasmid and chromosome segregation, and related systems, e.g.~those used to position chemotaxis protein clusters \cite{Roberts2012} and carboxysomes \cite{Savage2010}. 
These systems utilize ParA-like ATPases that bind in the ATP-bound form to the nucleoid and regulate the movement of a cargo (e.g.~protein cluster, partition complex or plasmids) to the required intracellular positions \cite{Howard2010}. 

Previously, we introduced a one-dimensional mathematical model of the dynamics of the PomXY cluster in \textit{M. xanthus} \cite{Schumacher2017a, Bergeler2018} that takes the elasticity of the nucleoid \cite{Wiggins2010, Lim2014} into account. 
Based on this model, we proposed a mechanism that relies on PomZ dimer fluxes on the nucleoid to generate the experimentally observed cluster dynamics, which is midcell localization. 
A flux-based mechanism was originally proposed by Ietswaart et al.~\cite{Ietswaart2014} for Par-mediated plasmid positioning. 
They showed that diffusive ParA fluxes can in principle explain equidistant spacing of plasmids along the nucleoid, because only when this is the case will the fluxes from either side of each plasmid balance up. 
Hence, macromolecular objects can be evenly distributed on the nucleoid if they move in the direction from which the larger number of ParA proteins impinges upon them \cite{Ietswaart2014}. 
In our model of the Pom system, an analogous mechanism can localize the PomXY cluster to mid-nucleoid \cite{Schumacher2017a, Bergeler2018}.

\begin{figure*}[t!]
\centering
\includegraphics{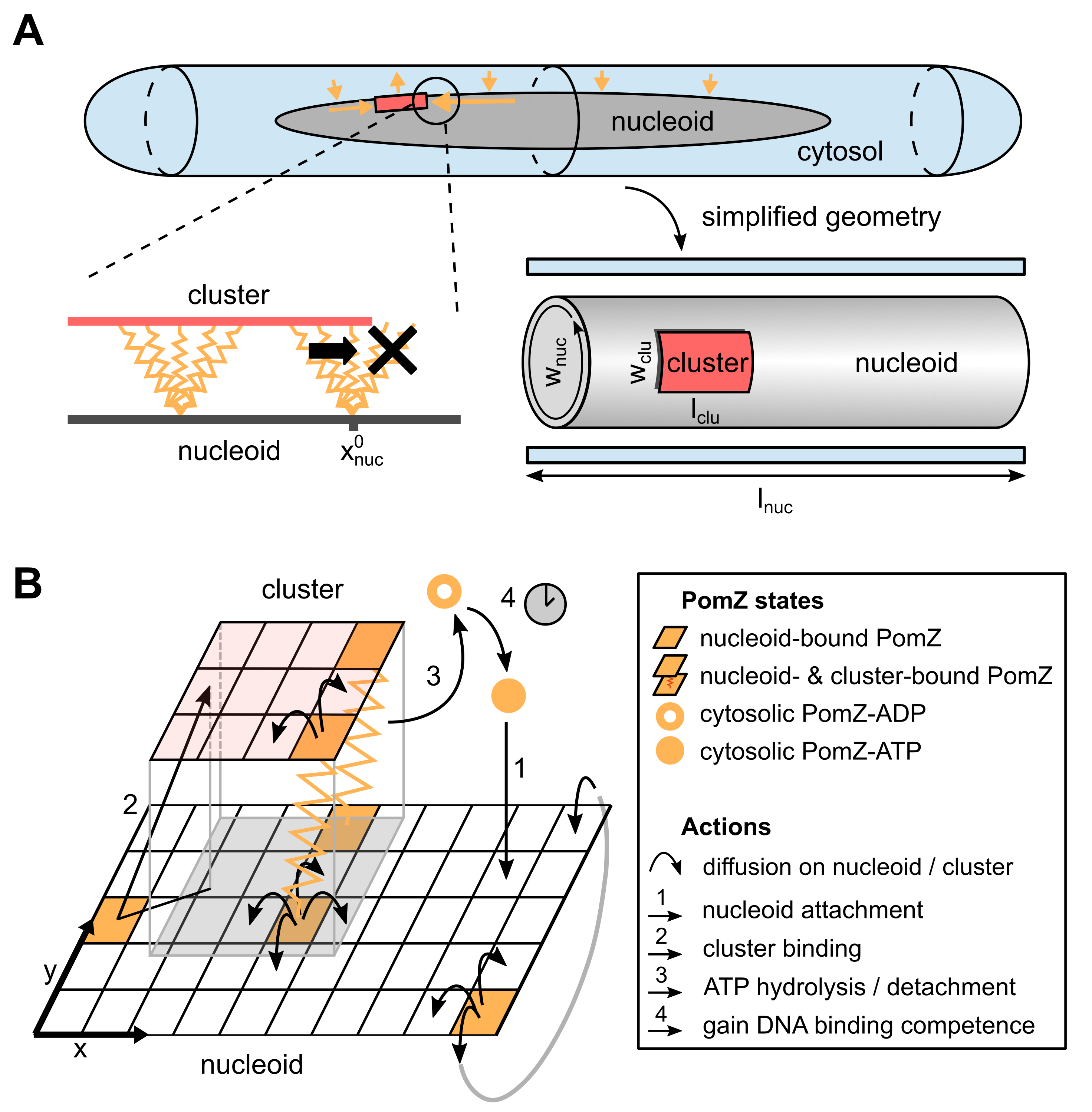}
\caption{\textbf{Mathematical model for PomXY cluster positioning, which accounts for the three-dimensional geometry of the cell.} 
(A)~Schematic representation of the geometry used in our model. 
Top: Sketch of a \textit{M. xanthus} cell. 
Cytosolic, ATP-bound PomZ can bind to the nucleoid (orange arrows towards the nucleoid), and diffuses on the nucleoid. 
When bound to the PomXY cluster, PomZ hydrolyzes ATP and is released into the cytosol (orange arrow away from the nucleoid). 
These dynamics lead to a net diffusive flux of PomZ on the nucleoid towards the cluster, which is larger from the side with the larger cluster-to-nucleoid end distance (horizontal orange arrows) \cite{Ietswaart2014}. 
Bottom left: 
PomZ dimers, modeled as springs, can exert forces on the cluster by binding to the cluster in a deflected configuration and by encountering the cluster's edge. 
For a particular nucleoid binding-site position, $\xnucz$, the positions available to PomZ's cluster-binding site are limited if the dimer is located close to the cluster's edge (black cross). 
This asymmetry can result in a force exerted on the cluster (black arrow). 
Bottom right: The model geometry derived from the biologically realistic three-dimensional cell. 
The nucleoid is modeled as an open cylinder and the cluster as an object of fixed size of rectangular shape. 
The cytosolic PomZ-ATP distribution is either assumed to be homogeneous or included effectively by modeling the cytosolic PomZ distribution along the long cell axis.
(B)~Schematic of the interactions of PomZ with the nucleoid and cluster considered in our model (see main text and SI text for details).}
\label{Fig_ModelSchematics2}
\end{figure*}

It remained unclear, however, whether such a flux-based mechanism can also localize midcell if the full three-dimensional geometry of the nucleoid is accounted for (see Figure~\ref{Fig_ModelSchematics2}A).
The reason is that, in contrast to the one-dimensional case, PomZ dimers can now more easily pass the cluster by without interacting with it. 
As a result, an asymmetry in the PomZ fluxes might be balanced. 
Here we show that the Pom cluster still localizes at midcell by a flux-based mechanism even if PomZ dimers can diffuse past the cluster.

\section{Results} 

\subsection{Flux-based mechanism for midcell localization}
\label{sec:flux-based_mechanism}

The dynamics of the PomXY cluster on the nucleoid crucially depends on the PomZ dynamics, as the cluster is tethered to the nucleoid via PomZ dimers \cite{Schumacher2017a}. 
Based on the biochemical processes suggested by experiments \cite{Schumacher2017a}, we model the dynamics of PomZ as follows (see Figure~\ref{Fig_ModelSchematics2}B and SI text for details).
ATP-bound PomZ dimers bind to the nucleoid with rate $k_\text{on}$ (action 1 in Figure~\ref{Fig_ModelSchematics2}B). 
Once on the nucleoid, they diffuse with diffusion constant $\Dnuc$.
The PomZ dimers are modeled effectively as springs to account for the elasticity of the chromosome and the Pom proteins (as in \cite{Lim2014, Surovtsev2016a}). 
For simplicity we will refer to the PomZ dimers as springs in the following, although the elasticity mainly originates from the nucleoid. 

A nucleoid-bound PomZ dimer can attach to the cluster with rate $k_a$ either `orthogonally', or obliquely in an extended configuration (action 2 in Figure~\ref{Fig_ModelSchematics2}B). 
We assume that cluster-bound PomZ can diffuse on both the nucleoid and the PomXY cluster.
However, the freedom of movement of nucleoid- and cluster-bound PomZ is restricted due to the energy cost involved in stretching the spring (for details see SI text).
Cluster-bound PomZ dimers remain attached to the cluster until they are released into the cytosol upon ATP hydrolysis, which is stimulated by PomX, PomY and DNA. 
ATP hydrolysis leads to a conformational change in the PomZ dimer and triggers the release of two ADP-bound PomZ monomers into the cytosol \cite{Schumacher2017a}. 
In our model, we combine these processes into one by using a single rate $k_h$ to describe the detachment of cluster-bound PomZ-ATP dimers into the cytosol as monomers (action 3 in Figure~\ref{Fig_ModelSchematics2}B). 
Before PomZ can rebind to the nucleoid, it must first bind ATP and dimerize. 
This introduces a delay between detachment from and reattachment to the nucleoid, which allows for spatial redistribution of the quickly diffusing cytosolic PomZ dimers in the cell (action 4 in Figure~\ref{Fig_ModelSchematics2}B). 

The cluster dynamics, which we approximate as over-damped, is determined by the forces exerted by the PomZ dimers on the cluster and the friction coefficient of the cluster (see SI text). 
Previously we suggested a flux-based mechanism for its midcell positioning in a one-dimensional model geometry \cite{Schumacher2017a, Bergeler2018}, which can be summarized as follows.
The PomZ dimers, which are modeled as springs, can exert net forces on the cluster in two different ways. 
They can attach to the cluster in a stretched configuration and thereby exert forces on the cluster (similar to the DNA-relay mechanism proposed for Par systems, \cite{Lim2014}). 
However, for PomZ dimers that quickly diffuse on the nucleoid (as observed experimentally in \textit{M. xanthus} cells) the initial deflection of the spring relaxes within a short time, such that only a small net force is exerted on the cluster. 
In this case, forces are mainly generated by cluster-bound PomZ dimers that encounter the cluster's edge  
(Figure~\ref{Fig_ModelSchematics2}A, bottom left) \cite{Bergeler2018}.
Every time a nucleoid- and cluster-bound PomZ dimer reaches the cluster's edge, the nucleoid binding site can diffuse beyond the cluster region, whereas the cluster binding site is restricted in its movement, which on average results in a net force on the cluster. 
These forces are such that a protein that reaches the cluster from the right exerts a force, which `drags' the cluster to the right and vice versa. 

The fact that PomZ dimers can exert forces on the cluster is not enough to explain the movement of the cluster towards midcell. 
Here, the asymmetry in the PomZ dimer density on the nucleoid, which prevails as long as the cluster is located off center, i.e.~not already at mid-nucleoid, is crucial. 
Since PomZ dimers can attach anywhere on the nucleoid but can detach only when they make contact with the cluster (as stimulation of their ATPase activity depends on PomX, PomY and DNA), a non-equilibrium flux of PomZ dimers in the system can be maintained. 
In particular, this flux includes a diffusive flux of PomZ dimers along the nucleoid towards the cluster. 
For a cluster located off-center the diffusive PomZ fluxes into the cluster from either side will differ \cite{Ietswaart2014}. 
Since the average forces applied by PomZ dimers arriving from the right and left sides of the cluster act in opposite directions, the difference in the PomZ fluxes from the two sides determines the net force exerted on the cluster \cite{Bergeler2018}. 
Taken together, these factors combine to drive a self-regulated midcell localization process as long as the PomZ dynamics is fast compared to the cluster dynamics and, if this is not the case, lead to oscillatory cluster movements along the nucleoid \cite{Bergeler2018}.

\subsection{A three-dimensional model for midcell localization}
To understand how the geometry of the nucleoid and the size of the PomXY cluster affect the cluster dynamics and thereby test whether a flux-based mechanism is feasible in a biologically realistic three-dimensional geometry, we investigated the mathematical model illustrated in Figure~\ref{Fig_ModelSchematics2}. 
Here, the nucleoid and the PomXY cluster are approximated as a cylindrical object and a rectangular sheet, respectively. 
Since experiments in \textit{M. xanthus} cells suggest that the cluster is large \cite{Schumacher2017a} we assume that the cluster, tethered to the nucleoid via PomZ dimers, moves over the nucleoid's surface and does not penetrate the bulk of the nucleoid.
Moreover, we assume that PomZ dimers also bind to and diffuse on the nucleoid's surface only.

The cylindrical geometry of the nucleoid is mathematically implemented by a rectangular sheet with periodic boundary conditions for the PomZ movements along the short cell axis ($y$ direction) and reflecting boundary conditions along the long cell axis ($x$ direction, Figure~\ref{Fig_ModelSchematics2}B). 
The cluster is modeled as a rectangle with reflecting boundaries at its edges for the PomZ dimer movement.
We refer to the extension of the cluster along the long and short cell axis as the cluster's length, $l_\mathrm{clu}$, and width, $w_\mathrm{clu}$, respectively (see Figure~\ref{Fig_ModelSchematics2}A). 

Besides the nucleoid and the cluster, the cytosol needs to be accounted for in our model, as PomZ dimers cycle between a nucleoid-bound and cytosolic state. 
We expect the cytosolic diffusion constant of PomZ to be of the same order as that of MinD proteins in \textit{E. coli} cells, $D_\text{cyt}\approx \SI{10}{\um^2 \second^{-1}}$ \cite{Meacci2006}. 
For ParA ATPases involved in chromosome and plasmid segregation, a delay between the release of ParA from the nucleoid into the cytosol and the re-acquisition of its capacity for non-specific DNA binding is observed experimentally \cite{Vecchiarelli2010}. 
Upon ATP hydrolysis and release of two ADP-bound ParA monomers into the cytosol, ParA must bind ATP, dimerize and then regain the competence to bind non-specifically to DNA before it can reattach to the nucleoid. 
The last step, a conformational change in the ATP-bound ParA dimer, is the slowest of these processes, occurring on a time scale of the order of minutes, $\tau \approx \SI{5}{\min}$ \cite{Vecchiarelli2010}. 
Since PomZ is ParA-like, we expect the PomZ ATPase cycle to be similar to the ParA cycle, and therefore assume the corresponding reaction rates to be of the same order of magnitude. 
This yields the following estimate for the diffusive length of cytosolic PomZ, $L_\text{diff}=\sqrt{D_\text{cyt}\tau}\approx\SI{55}{\um}$, which is significantly larger than the average cell length of a \textit{M. xanthus} cell of $\SI{7,7}{\um}$ \cite{Schumacher2017a}.
Hence, the assumption of a well-mixed PomZ density in the cytosol is justified \cite{Thalmeier2016}. 

However, it is not known how the cytosolic distribution of PomZ affects the cluster dynamics in our proposed flux-based mechanism. 
In particular, does a realistic, non-uniform distribution increase or reduce the speed of the cluster movement towards midcell? 
To obtain a qualitative answer to this question, we accounted for the cytosolic PomZ distribution in a simplified way by focusing on the variation in PomZ density along the long cell axis and approximating the density along the short cell axis as uniform (Figure~\ref{Fig_ModelSchematics2}A, bottom right). 
In section \ref{sec:cytosol_distribution} we discuss how the cluster dynamics is affected when PomZ's diffusion constant in the cytosol is reduced (which leads to deviations from the uniform PomZ distribution), but for now we assume a homogeneous PomZ-ATP distribution in the cytosol. 

\begin{figure*}[t!]
\centering
\includegraphics[width = 0.8\textwidth]{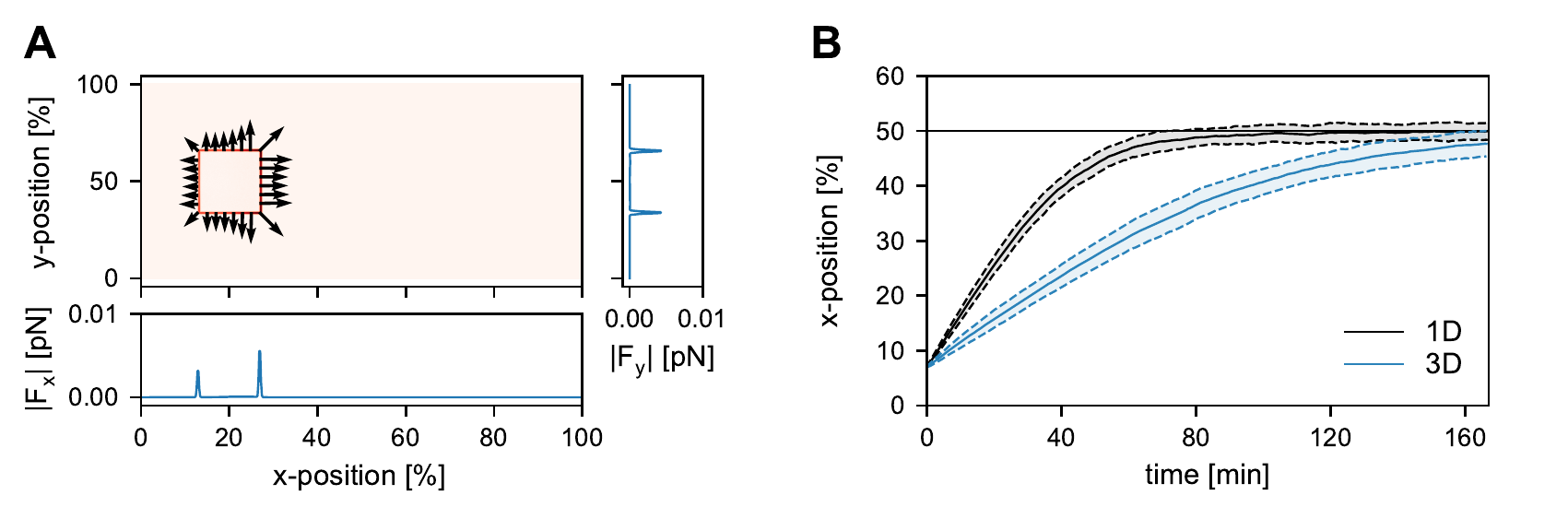}
\caption{\textbf{A flux-based mechanism can explain midcell localization in a three-dimensional cell geometry.} 
(A)~For a fixed cluster position (here at 20\% nucleoid length), the average forces exerted by PomZ dimers on the cluster are plotted per nucleoid lattice site.
The color code shows the magnitude of the average force vector, which is highest at the cluster's edges (darker red indicates higher values). 
At these edges, the average force vectors are plotted, $\vec{F} =(F_x, F_y)$. 
The average $x$- and $y$-component of the force, $F_x$ and $F_y$, (summed over all $y$- and $x$-positions, respectively), are shown in the lower and right panels, respectively. 
(B)~Comparison of the average cluster trajectories along the cell's long axis, obtained from the three-dimensional model and its one-dimensional counterpart studied previously \cite{Bergeler2018} (see also Figure S3). 
We averaged over an ensemble of $100$ simulations. 
The shaded regions depict one standard deviation around the mean density value. 
In all cases, the cluster is initially positioned at the left edge of the nucleoid such that it overlaps entirely with the nucleoid (at $7\%$ of the nucleoid length). 
Mid-nucleoid is indicated by the horizontal black line. 
The parameter values used in the simulations are given in Table S1. 
}
\label{Fig_3dResults}
\end{figure*} 

\subsection{A flux-based mechanism can explain midcell positioning in three dimensions}
\label{sec:flux_based_model}
Our simulations using the three-dimensional model geometry show that the net force exerted by the PomZ dimers on a stationary cluster (i.e.~a cluster with fixed position) is still directed towards mid-nucleoid (Figure~\ref{Fig_3dResults}A). 
The asymmetry in the forces can be attributed to an asymmetry in the PomZ fluxes along the nucleoid into the cluster. 
This finding indicates that the cluster movement is biased towards midcell in the three-dimensional model geometry also.
Hence, a flux-based mechanism is also conceivable in three dimensions. 
Indeed, the simulated PomXY cluster trajectories show movement towards and localization at mid-nucleoid (Figure~\ref{Fig_3dResults}B). 
If the cluster's width does not cover the complete nucleoid circumference, the movement towards midcell takes longer than in the one-dimensional case (Figure \ref{Fig_3dResults}B).
The forces along the short cell axis direction balance (Figure \ref{Fig_3dResults}A) and hence no bias in the cluster movement along the short axis is expected for a slowly moving cluster and fast PomZ dynamics. 
The simulated cluster trajectories show diffusive motion along the short cell axis direction for the parameters given in Table S1 (see SI text). 

\subsection{\label{sec:cluster_size} The cluster's linear dimensions determine the time taken to reach midcell}

\begin{figure*}[tp!]
\centering
\includegraphics[width=0.85\textwidth]{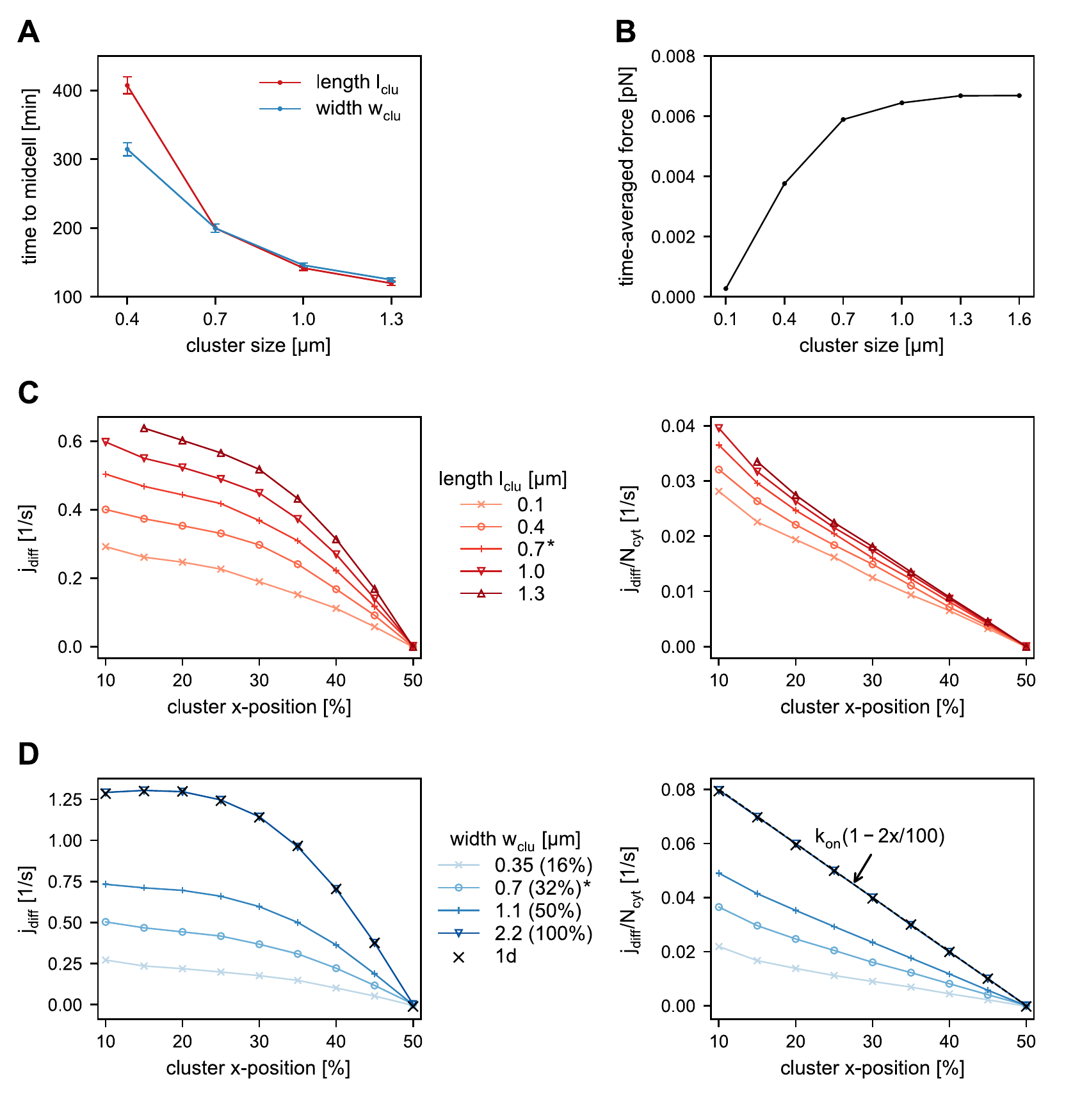}
\caption{\textbf{Dependence of the time needed to reach midcell on the size of the cluster.} 
(A)~Average first passage time of the PomXY cluster to reach mid-nucleoid for different cluster sizes. 
In all simulations the cluster starts at $13\%$ of nucleoid length, which corresponds to the leftmost position possible such that for all cluster sizes considered a full overlap with the nucleoid is ensured. 
The error bars show the standard error of the mean. 
(B)~Ensemble average of the time-averaged force exerted by a single PomZ dimer on the cluster for a one-dimensional model geometry and different cluster lengths. 
(C)-(D)~The PomZ flux difference into the cluster, $j_\text{diff}$, along the long cell axis for a cluster at a fixed position, which is varied from $10\%$ to $50\%$ of nucleoid length, is shown (see also Figure S2).
In (C), the cluster's length is varied and in (D) the cluster's width. 
For a ring-shaped cluster, $\wclu = \wnuc$, the simulation results agree with those from the one-dimensional model (black crosses). 
To understand how the cluster's length and width affect the flux difference, the values are scaled with the number of PomZ dimers in the cytosol, $N_\text{cyt}$, for each cluster position separately (Figures on the right). 
In (D) an analytical estimate for the flux difference is plotted (dashed line), which agrees with the simulation results for a ring-shaped cluster (see SI text for details). 
If not given explicitly, the parameter values used are those listed in Table S1 (data marked with a star).} 
\label{Fig_ClusterShape}
\end{figure*}

In this section we ask how the arrival time of the cluster at mid-nucleoid depends on the linear dimensions of the cluster.  
Our simulations show that the larger the cluster's length or width, the faster the cluster moves towards midcell (Figure \ref{Fig_ClusterShape}A).
In the following we discuss how these two observations can be explained heuristically.

The net force applied to the cluster by PomZ dimers depends on the average force exerted by a single PomZ dimer and on the flux difference in PomZ dimers arriving at the cluster \cite{Bergeler2018}. 
The latter can be regarded as the frequency of PomZ interactions with the cluster that lead to a net force contribution.  

The flux difference into the cluster increases for larger or broader clusters for two reasons.
First, increasing the length or width of the cluster results in an increased number of cluster-bound PomZ dimers, as the cluster becomes more accessible to nucleoid-bound PomZ dimers. 
This increase in cluster-bound PomZ dimers leads to a larger turnover of PomZ dimers cycling between the nucleoid-bound and cytosolic state, which also increases the flux difference. 
Second, the larger and wider the cluster, the less likely it is that PomZ dimers will diffuse past it without attaching, which would otherwise reduce the flux difference into the cluster along the long cell axis.

The forces exerted by PomZ dimers also depend on the linear dimensions of the cluster. 
Increasing the cluster size increases the average force exerted by a single PomZ dimer, until a maximal force is reached (Figure~\ref{Fig_ClusterShape}B). 
This dependence can be explained by diffusion of PomZ dimers bound to the cluster:
The smaller the cluster size, the more likely it is that a PomZ dimer attaching close to one edge of the cluster will reach the other edge before detaching into the cytosol.
Since the PomZ dimer then exerts forces in both directions along the long cell axis, the time-averaged force it applies to the cluster is reduced.

Based on these observations, we can rationalize the dependence of the arrival time at mid-nucleoid on the length of the cluster as follows.
An increase in cluster length (while keeping its width constant) increases the numbers of PomZ dimers interacting with the cluster, and hence the flux difference along the long cell axis. 
Indeed, our simulation results show that the longer the cluster, the larger the difference in the PomZ dimer fluxes from either side along the direction of the long cell axis (Figure~\ref{Fig_ClusterShape}C).
To test whether the increased flux of PomZ dimers in the system is the main determinant for the observed increase in the flux difference, we scaled the fluxes by the number of PomZ dimers in the cytosol, $N_\text{cyt}$, which is proportional to the flux of PomZ dimers onto the nucleoid. 
Upon rescaling, the flux differences decay approximately linearly with the cluster position (Figure~\ref{Fig_ClusterShape}C, right). 
However, longer clusters still show the largest flux differences. 
This phenomenon can be attributed to the fact that for shorter clusters PomZ dimers are more likely to diffuse past the cluster.

For a longer cluster not only the diffusive flux of PomZ dimers into the cluster, but also the force exerted by a single PomZ dimer on the cluster is increased (Figure~\ref{Fig_ClusterShape}B). 
Hence, frequency and magnitude of forces exerted on the cluster are increased, implying a larger net force, which explains the shorter arrival times of the cluster at mid-nucleoid (Figure~\ref{Fig_ClusterShape}A). 

Next, we discuss the dependence of the arrival time on the cluster width. 
As in the case of an increase in length, the overall turnover of PomZ dimers increases with cluster width. 
The width of the cluster determines how many PomZ dimers approach and attach to the cluster from the directions of the long and the short cell axes, respectively, and how many pass the cluster without interacting with it.
The broader the cluster, the smaller the flux of PomZ dimers that can diffuse past the cluster without attaching. 
Our simulation results show an increased PomZ flux difference when the cluster width is increased from $16\%$ to $100\%$ of the nucleoid's circumference (Figure~\ref{Fig_ClusterShape}D).
When the cluster covers the entire circumference of the nucleoid, we call it a ring.
Rescaling of the flux differences with the number of PomZ dimers in the cytosol, $N_\text{cyt}$, leads to values that are still larger for wider clusters, due to the reduced flux of PomZ dimers past the cluster (Figure~\ref{Fig_ClusterShape}D, right). 
The forces exerted by single PomZ dimers in the direction of the long cell axis direction are not affected by a change in the cluster width.
Hence, the decrease in arrival time for wider clusters can be explained by the increased PomZ flux difference into the cluster along the long cell axis alone.

\subsection{\label{sec:cytosol_distribution} Cytosolic diffusion ensures fast midcell positioning of the cluster}

\begin{figure}[!]
\centering
\includegraphics[width = 0.84\columnwidth]{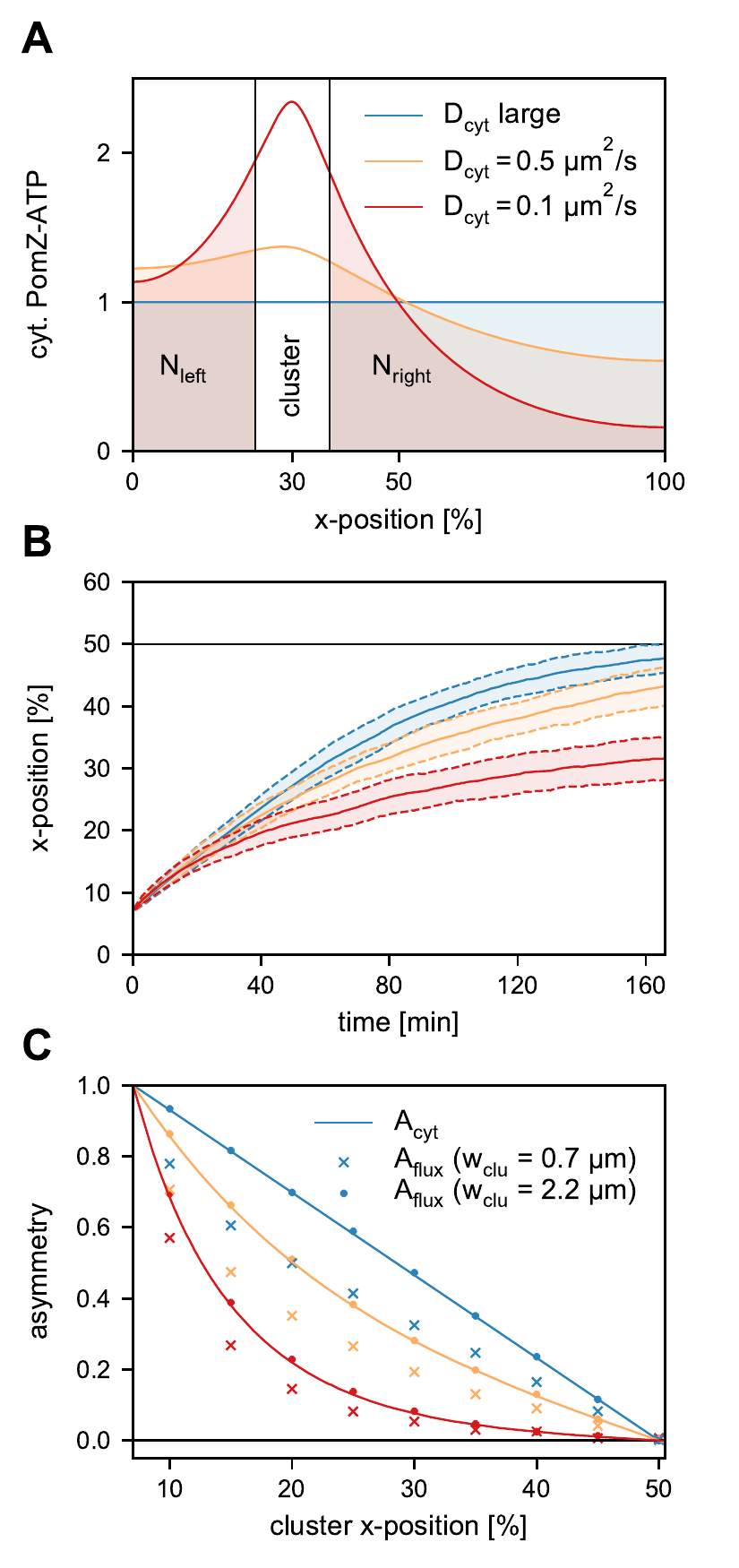}
\caption{\textbf{Fast cytosolic diffusion accelerates midcell localization.} 
(A)~Cytosolic PomZ-ATP distribution along the long axis ($x$ axis) for different cytosolic PomZ diffusion constants $D_\text{cyt}$ (see also Figure S1). 
Integrating the distributions along the long cell axis left and right of the cluster yields the total number of PomZ-ATP proteins left and right of the cluster, $\Nleft$ and $\Nright$. 
(B)~Average cluster trajectories along the $x$ direction for the different cytosolic PomZ diffusion constants in (A). 
The shading denotes the regions of one standard deviation around the average trajectories. 
Mid-nucleoid is indicated by the solid black line. In the simulations, the clusters are positioned initially such that the left edge of the cluster coincides with the left edge of the nucleoid. 
(C)~Asymmetry measure of the number of cytosolic PomZ-ATP left and right of the cluster, $A_\text{cyt}$ (Equation~\ref{Eq_asymm}) (solid lines), compared to the PomZ flux asymmetry into the cluster, $A_\text{flux}$ (Equation~\ref{eq:Aflux}), for different cytosolic PomZ diffusion constants.
The dots indicate the flux asymmetry into a cluster that forms a ring. 
Crosses indicate the asymmetry for a cluster with a width of $32\%$ of the nucleoid's circumference (same value as in Table S1). We averaged over $100$ runs of the simulation. 
The parameter values given in Table S1 are used if not explicitly stated otherwise.} 
\label{Fig_CytosolicDiffusion}
\end{figure}

So far, we have assumed that the cytosolic PomZ distribution is spatially uniform. 
Now we investigate how the cluster dynamics change when spatial heterogeneity in the cytosolic PomZ distribution is included in the model. 
To this end, we explicitly incorporate the cytosol by approximating the cytosolic volume as a one-dimensional layer of the same length as the nucleoid (see Figure~\ref{Fig_ModelSchematics2}A) and formulate reaction-diffusion equations for the density of cytosolic PomZ-ADP ($c_D$) and PomZ-ATP ($c_T$) along the long cell axis. 
For simplicity, we consider only an active (ATP-bound) and inactive (ADP-bound) conformation of PomZ, and disregard any explicit monomeric and dimeric states of PomZ in the cytosol (for details see SI text).
The stationary solution for $c_T(x; x_c)$ deviates most from a uniform distribution, the smaller the cytosolic diffusion constant (Figure \ref{Fig_CytosolicDiffusion}A).
In the limit of infinitely large cytosolic PomZ diffusion constants, the cytosolic PomZ distribution becomes spatially uniform. 

To investigate the effect of the cytosolic PomZ distribution on the cluster's trajectory, we replaced the spatially uniform cytosolic PomZ-ATP distribution by $c_T(x; x_c)$, in our three-dimensional model.
Since \textit{M. xanthus} cells are rod-shaped with the length being much larger than the width, we approximate the cytosolic PomZ-ATP distribution along the short cell axis as uniform.
With this spatially heterogeneous attachment rate to the nucleoid, our simulations show that for a larger deviation of the cytosolic PomZ distribution from a spatially uniform one (decreasing $D_\text{cyt}$), the movement of the clusters is less biased towards mid-nucleoid (Figure~\ref{Fig_CytosolicDiffusion}B). 
We conclude that the cytosolic PomZ distribution has an impact on the cluster trajectories and the velocity of the cluster towards mid-nucleoid is maximal for a uniform cytosolic PomZ-ATP distribution.

How does the cytosolic PomZ distribution affect the cluster's movement?
The flux of PomZ dimers onto the nucleoid region to the left of the cluster and the diffusive flux of PomZ dimers into the cluster from the left are equal for a stationary cluster in the one-dimensional model geometry, and vice versa for the right side (see SI text).
Since the total flux of PomZ dimers onto the nucleoid to the left and right of the cluster depends on the cytosolic PomZ-ATP distribution, we expect the diffusive PomZ fluxes into the cluster to depend on this distribution as well. 
To investigate this further, we define the following asymmetry quantity
\begin{equation}
A_\text{cyt}=\frac{N_\text{right}-N_\text{left}}{N_\text{right}+N_\text{left}},
\label{Eq_asymm}
\end{equation}
which compares the total numbers of cytosolic PomZ-ATP left and right of the cluster (see Figure \ref{Fig_CytosolicDiffusion}A): 
\begin{align}
N_\text{left} &= \int_0^{x_c - \lclu/2} c_T(x; x_c) \mathrm{d} x~, \\
N_\text{right} &= \int_{x_c + \lclu/2}^{\lnuc} c_T(x; x_c) \mathrm{d} x~.
\end{align}
The corresponding asymmetry in the PomZ fluxes into the cluster from the left, $j_\text{left}$, and right, $j_\text{right}$, sides is given by
\begin{equation}
\label{eq:Aflux}
A_\text{flux} =  \frac{j_\text{right}-j_\text{left}}{j_\text{right}+j_\text{left}}~.
\end{equation} 

We measured this flux asymmetry in our simulations for two scenarios. 
First, for a PomXY cluster that forms a ring around the nucleoid and second, for a cluster that does not entirely encompass the nucleoid's circumference. 
We find that the asymmetry in the flux, $A_\text{flux}$, obtained from simulations for a cluster that forms a ring, and the corresponding asymmetry in the cytosolic PomZ-ATP density, $A_\text{cyt}$, agree nicely (Figure~\ref{Fig_CytosolicDiffusion}C).
Hence, an asymmetry in the cytosolic distribution of PomZ-ATP is directly reflected in the diffusive PomZ fluxes into the cluster. 
For an infinitely large cytosolic PomZ diffusion constant, both asymmetry measures decay linearly when the cluster position is varied from far off-center towards mid-nucleoid. 
Decreasing the cytosolic PomZ diffusion constant results in asymmetry curves that decay faster than linearly towards zero (Figure~\ref{Fig_CytosolicDiffusion}C). 
For a cluster that does not cover the whole nucleoid's circumference, the asymmetry in the PomZ fluxes into the cluster is smaller than the asymmetry in the cytosolic PomZ-ATP concentration (Figure~\ref{Fig_CytosolicDiffusion}C). 
This can be attributed to the reduction in the diffusive fluxes of PomZ dimers into the cluster along the long cell axis, as discussed before. 
The reduced asymmetry in the diffusive PomZ fluxes explains the less biased movement of the cluster towards mid-nucleoid for smaller cytosolic PomZ diffusion constants (Figure \ref{Fig_CytosolicDiffusion}B).

\subsection{Two clusters localize at one- and three-quarter positions}
Motivated by equidistant positioning of multiple cargoes, such as plasmids \cite{Ringgaard2009, Ebersbach2006, Sengupta2010, Ietswaart2014, Hatano2010}, we also considered the dynamics of two PomXY clusters in the three-dimensional model geometry.
Our simulation results show equidistant positioning of the two clusters (see SI text, Figure S4). 

\section{Discussion}
\label{sec:discussion}
In this work we investigated a mathematical model for midcell localization in \textit{M. xanthus} using a biologically realistic three-dimensional geometry for the nucleoid. 
Whether or not a flux-based mechanism can position macromolecular objects when the ATPases (here PomZ) can diffuse past a cargo (here PomXY cluster) has been questioned, because the fluxes into the cargo might equalize \cite{Ietswaart2014}. 
We showed that if PomZ dimers can diffuse past the PomXY cluster, there is still a flux difference into the cluster (for a cluster positioned off-center), which leads to a bias in the cluster movement towards mid-nucleoid. 
Hence, we conclude that a flux-based mechanism can explain midcell positioning of one or equidistant localization of several Pom clusters even if PomZ can diffuse past the cluster.  

To investigate the effect of the flux of PomZ dimers past the cluster, we studied the dependence of cluster dynamics on its width and length. 
We find that increasing the cluster length or width shortens the time taken for the cluster to reach mid-nucleoid.
This can be attributed to an overall increase in the flux difference, reduced flux past the cluster and the larger forces single PomZ dimers exert, on average, on larger and wider clusters. 

Our simulation data further demonstrates that fast cytosolic diffusion of PomZ proteins reduces the time taken to find midcell. 
This finding is in agreement with previous results for dynamic protein clusters in bacterial cells \cite{Murray2017}. 
The asymmetry in the cytosolic PomZ density left and right of the cluster along the long cell axis is reflected in the diffusive flux difference of PomZ dimers into the cluster, and thus influences the cluster dynamics. 
If PomZ diffuses quickly in the cytosol, the cytosolic PomZ-ATP distribution becomes spatially uniform. 
In this case, the flux of PomZ dimers onto the nucleoid scales with the length of the nucleoid regions left and right of the cluster. 
This results in the largest PomZ flux differences (for an off-center cluster) compared to spatially non-uniform cytosolic PomZ-ATP distributions.  
Interestingly, spatial redistribution of proteins in the cytosol is also found to be important for other pattern-forming systems, including Min protein pattern formation \cite{Halatek2018, Halatek2018b} and ParA-mediated cargo movement \cite{Vecchiarelli2010}. 

Le Gall et al.~\cite{LeGall2016} have shown that partition complexes as well as plasmids move within the nucleoid volume. 
In contrast, based on the large size of the PomXY cluster, we assumed that the movement of the cluster, tethered to the nucleoid via PomZ dimers, is restricted to the surface of the nucleoid.
To verify our assumption, the position of the cluster relative to the nucleoid needs to be measured \textit{in vivo} using e.g.\ super-resolution microscopy.
In addition, since PomZ dimers are much smaller than the PomXY cluster, they might be able to diffuse within the nucleoid volume even though the cluster does not.
It would be interesting to investigate this aspect further. 

In summary, we have shown that a flux-based mechanism can explain midcell localization of one, and equidistant positioning of two clusters in a model geometry that allows the ATPase PomZ to diffuse past the clusters on the nucleoid.
This observation is also important for other positioning systems, such as the Par system, which equidistantly spaces low-copy-number plasmids along the nucleoid. 
Understanding the differences and similarities between these positioning systems will help us to understand the generic mechanisms underlying the localization patterns of cargoes inside the cell. 

\section*{Acknowledgements}
The authors thank Isabella Graf, Emanuel Reithmann, Christoph Brand, Dominik Schumacher and Lotte S{\o}gaard-Andersen for helpful discussions. 
This research was supported by a DFG fellowship through the Graduate School of Quantitative Biosciences Munich, QBM (SB, EF), the Deutsche Forschungsgemeinschaft (DFG) via project P03 within the Transregio Collaborative Research Center (TRR 174) ``Spatiotemporal Dynamics of Bacterial Cells'' (SB, EF), and the German Excellence Initiative via the program ``Nanosystems Initiative Munich'' (EF).


\begin{thebibliography}{10}

\bibitem{Gerdes2010}
K.~Gerdes, M.~Howard, and F.~Szardenings, ``{Pushing and pulling in prokaryotic
  DNA segregation},'' {\em Cell}, vol.~141, no.~6, pp.~927--942, 2010.

\bibitem{Lutkenhaus2012}
J.~Lutkenhaus, ``{The ParA/MinD family puts things in their place},'' {\em
  Trends Microbiol}, vol.~20, pp.~411--418, 2012.

\bibitem{Thalmeier2016}
D.~Thalmeier, J.~Halatek, and E.~Frey, ``{Geometry-induced protein pattern
  formation},'' {\em Proc Natl Acad Sci U S A},
  vol.~113, no.~3, pp.~548--553, 2016.

\bibitem{Schumacher2017a}
D.~Schumacher, S.~Bergeler, A.~Harms, J.~Vonck, S.~Huneke-Vogt, E.~Frey, and
  L.~S{\o}gaard-Andersen, ``{The PomXYZ proteins self-organize on the bacterial
  nucleoid to stimulate cell division.},'' {\em Dev Cell}, vol.~41,
  pp.~299--314, 2017.

\bibitem{Treuner-Lange2013}
A.~Treuner-Lange, K.~Aguiluz, C.~van~der Does, N.~G{\'{o}}mez-Santos, A.~Harms,
  D.~Schumacher, P.~Lenz, M.~Hoppert, J.~Kahnt, J.~Mu{\~{n}}oz-Dorado, and
  L.~S{\o}gaard-Andersen, ``{PomZ, a ParA-like protein, regulates Z-ring
  formation and cell division in \textit{Myxococcus xanthus}},'' {\em Mol Microbiol},
  vol.~87, pp.~235--253, 2013.

\bibitem{Schumacher2017}
D.~Schumacher and L.~S{\o}gaard-Andersen, ``{Regulation of cell polarity in
  motility and cell division in \textit{Myxococcus xanthus}},'' {\em Annu Rev
  Microbiol}, vol.~71, pp.~61--78, 2017.

\bibitem{Roberts2012}
M.~A.~J. Roberts, G.~H. Wadhams, K.~A. Hadfield, S.~Tickner, and J.~P.
  Armitage, ``{ParA-like protein uses nonspecific chromosomal DNA binding to
  partition protein complexes},'' {\em Proc Natl Acad Sci U S A}, vol.~109, no.~17, pp.~6698--6703, 2012.

\bibitem{Savage2010}
D.~F. Savage, B.~Afonso, A.~H. Chen, and P.~A. Silver, ``{Spatially ordered
  dynamics of the bacterial carbon fixation machinery},'' {\em Science},
  vol.~327, no.~5970, pp.~1258--1261, 2010.

\bibitem{Howard2010}
M.~Howard and K.~Gerdes, ``{What is the mechanism of ParA-mediated DNA
  movement?},'' {\em Mol Microbiol}, vol.~78, pp.~9--12, 2010.
  
\bibitem{Bergeler2018}
S.~Bergeler and E.~Frey, ``{Regulation of Pom cluster dynamics in \textit{Myxococcus
  xanthus}},'' {\em PLoS Comput Biol}, vol.~14, no.~8, p.~e1006358, 2018.
  
\bibitem{Wiggins2010}
P.~A. Wiggins, K.~C. Cheveralls, J.~S. Martin, R.~Lintner, and J.~Kondev,
  ``{Strong intranucleoid interactions organize the Escherichia coli chromosome
  into a nucleoid filament},'' {\em Proc Natl Acad Sci U S A}, vol.~107, no.~11, pp.~4991--4995, 2010.

\bibitem{Lim2014}
H.~C. Lim, I.~V. Surovtsev, B.~G. Beltran, F.~Huang, J.~Bewersdorf, and
  C.~Jacobs-Wagner, ``{Evidence for a DNA-relay mechanism in ParABS-mediated
  chromosome segregation},'' {\em Elife}, vol.~3, p.~e02758, 2014.

\bibitem{Ietswaart2014}
R.~Ietswaart, F.~Szardenings, K.~Gerdes, and M.~Howard, ``{Competing ParA
  structures space bacterial plasmids equally over the nucleoid},'' {\em PLoS
  Comput Biol}, vol.~10, no.~12, 2014.

\bibitem{Surovtsev2016a}
I.~V. Surovtsev, M.~Campos, and C.~Jacobs-Wagner, ``{DNA-relay mechanism is
  sufficient to explain ParA-dependent intracellular transport and patterning
  of single and multiple cargos},'' {\em Proc Natl Acad Sci U S A}, vol.~113, no.~46,
  pp.~E7268--E7276, 2016.
  
\bibitem{Meacci2006}
G.~Meacci, J.~Ries, E.~Fischer-Friedrich, N.~Kahya, P.~Schwille, and K.~Kruse,
  ``{Mobility of Min-proteins in \textit{Escherichia coli} measured by fluorescence
  correlation spectroscopy},'' {\em Phys Biol}, vol.~3, no.~4,
  pp.~255--263, 2006.
  
\bibitem{Vecchiarelli2010}
A.~G. Vecchiarelli, Y.-W. Han, X.~Tan, M.~Mizuuchi, R.~Ghirlando,
  C.~Biert{\"{u}}mpfel, B.~E. Funnell, and K.~Mizuuchi, ``{ATP control of
  dynamic P1 ParA-DNA interactions: a key role for the nucleoid in plasmid
  partition},'' {\em Mol Microbiol}, vol.~78, pp.~78--91, 2010.
  
\bibitem{Ringgaard2009}
S.~Ringgaard, J.~van Zon, M.~Howard, and K.~Gerdes, ``{Movement and
  equipositioning of plasmids by ParA filament disassembly},'' {\em Proc Natl Acad Sci U S A},
  vol.~106, pp.~19369--19374, 2009.
  
\bibitem{Ebersbach2006}
G.~Ebersbach, S.~Ringgaard, J.~M{\o}ller-Jensen, Q.~Wang, D.~J. Sherratt, and
  K.~Gerdes, ``{Regular cellular distribution of plasmids by oscillating and
  filament-forming ParA ATPase of plasmid pB171},'' {\em Mol
  Microbiol}, vol.~61, no.~6, pp.~1428--1442, 2006.

\bibitem{Sengupta2010}
M.~Sengupta, H.~J. Nielsen, B.~Youngren, and S.~Austin, ``{P1 plasmid
  segregation: Accurate redistribution by dynamic plasmid pairing and
  separation},'' {\em J Bacteriol}, vol.~192, no.~5,
  pp.~1175--1183, 2010.

\bibitem{Hatano2010}
T.~Hatano and H.~Niki, ``{Partitioning of P1 plasmids by gradual distribution
  of the ATPase ParA},'' {\em Mol Microbiol}, vol.~78, no.~5,
  pp.~1182--1198, 2010.

\bibitem{Murray2017}
S.~M. Murray and V.~Sourjik, ``{Self-organization and positioning of bacterial
  protein clusters},'' {\em Nat Phys}, vol.~13, pp.~1006--1013, 2017.
  
\bibitem{Halatek2018}
J.~Halatek and E.~Frey, ``{Rethinking pattern formation in reaction-diffusion
  systems},'' {\em Nat Phys}, 2018.
  
\bibitem{Halatek2018b}
J.~Halatek, F.~Brauns, and E.~Frey, ``{Self-organization principles of
  intracellular pattern formation},'' {\em Philos Trans R Soc Lond B Biol Sci}, vol.~373, p.~20170107, 2018.

\bibitem{LeGall2016}
A.~{Le Gall}, D.~I. Cattoni, B.~Guilhas, C.~Mathieu-Demazi{\`{e}}re,
  L.~Oudjedi, J.-B. Fiche, J.~Rech, S.~Abrahamsson, H.~Murray, J.-Y. Bouet, and
  M.~Nollmann, ``{Bacterial partition complexes segregate within the volume of
  the nucleoid},'' {\em Nat Commun}, vol.~7, p.~12107, 2016.

\bibitem{Lansky2015}
Z.~Lansky, M.~Braun, A.~L{\"{u}}decke, M.~Schlierf, P.~R.~ten Wolde, M.~E.~Janson, and S.~Diez, ``{Diffusible crosslinkers generate directed forces in microtubule networks}," {\em Cell}, vol.~160, pp.~1159--1168, 2015.

\bibitem{Hu2017a} 
L.~Hu, A.~G. Vecchiarelli, K.~Mizuuchi, K.~C. Neuman, and J.~Liu, ``{Brownian
  ratchet mechanism for faithful segregation of low-copy-number plasmids},''
  {\em Biophys J}, vol.~112, no.~7, pp.~1489--1502, 2017.
  
\bibitem{Gillespie2007}
D.~T. Gillespie, ``{Stochastic simulation of chemical kinetics},'' {\em Annu Rev Phys Chem}, vol.~58, pp.~35--55, 2007.

\bibitem{Hwang2013}
L.~C. Hwang, A.~G. Vecchiarelli, Y.-W. Han, M.~Mizuuchi, Y.~Harada, B.~E.
  Funnell, and K.~Mizuuchi, ``{ParA-mediated plasmid partition driven by
  protein pattern self-organization.},'' {\em EMBO J}, vol.~32,
  pp.~1238--1249, 2013.
  
\bibitem{Vecchiarelli2014}
A.~G. Vecchiarelli, K.~C. Neuman, and K.~Mizuuchi, ``{A propagating ATPase
  gradient drives transport of surface-confined cellular cargo},'' {\em
  Proc Natl Acad Sci U S A}, vol.~111, no.~13, pp.~4880--4885, 2014.
  
\bibitem{Jindal2015} 
L.~Jindal and E.~Emberly, ``{Operational principles for the dynamics of the in
  vitro ParA-ParB system},'' {\em PLoS Comput Biol}, vol.~11,
  p.~e1004651, 2015.

\end{thebibliography}
\end{document}


\begin{center}
\textbf{\large Supplemental Material}
\end{center}

\setcounter{equation}{0}
\setcounter{figure}{0}
\setcounter{table}{0}
\setcounter{page}{1}
\setcounter{section}{0}
\makeatletter
\renewcommand{\theequation}{S\arabic{equation}}
\renewcommand{\thefigure}{S\arabic{figure}}
\renewcommand{\thetable}{S\arabic{table}}
\renewcommand{\thesubsection}{\arabic{subsection}}

\section{\label{sec:SI_model}Details on the mathematical model}

\subsection{Attachment rate of PomZ to the nucleoid}
In our simulations we either assume that the PomZ density in the cytosol is homogeneous such that the flux of PomZ dimers onto the nucleoid is constant along the nucleoid or we account for the cytosolic PomZ distribution in a simplified manner. 
In the former case, a cytosolic PomZ dimer attaches to each lattice site of the nucleoid, also where the cluster is located, with the same rate. 
This rate is given by the attachment rate to the entire surface of the nucleoid, $k_\text{on}$, divided by the number of lattice sites $\lnuc \cdot \wnuc/a^2$. 
By $a$ we denote the lattice spacing, which has the same value in $x$- and $y$-direction. 
In the latter case, when we account for the cytosolic PomZ distribution, we replace the homogeneous with the steady-state cytosolic PomZ-ATP distribution in our model (see section \ref{sec:SI_cyt_PomZ}).
Here, we use a one-dimensional solution for the cytosolic PomZ-ATP density, which describes the variation of the protein density along the long cell axis. 
In our model, which has a three-dimensional geometry, we assume that the cytosolic PomZ-ATP density is uniform along the short cell axis and changes along the long cell axis according to the one-dimensional distribution. 

\subsection{Attachment of PomZ to and detachment from the PomXY cluster}
A nucleoid-bound PomZ dimer can bind to a lattice site on the PomXY cluster.  
Since PomZ dimers are modelled as springs to account for the elasticity of the nucleoid, they can attach to the cluster not only `orthogonally', but also in a stretched configuration.  
We approximate the rate for a PomZ dimer, bound to the nucleoid at position $\xnuc$, to attach to the cluster at position $\xclu$, as follows:
\begin{equation}
\label{Eq_Sup_RateDoubleAttachment}
k_a(\vec{x}_\text{clu}) = k_{a}^0 \cdot \exp{\left[- \frac 1 2 \, \beta k (\vec{x}_\text{clu}-\vec{x}_\text{nuc})^{2}\right]}~.
\end{equation}
The rate for attachment to a lattice site at position $\xclu$ on the cluster is then given by $k_a(\vec{x}_\text{clu}) \cdot a^2$. 
In the formula above, we multiply the constant rate $k_{a}^0$ with a Boltzmann factor corresponding to the distribution of the elongation of a spring in a thermal heat bath with temperature $T$. 
Here, $k$ denotes the effective spring stiffness of the PomZ dimers and $\beta$ is the inverse of the thermal energy, $\beta = 1/k_BT$. 
The positions of the cluster and nucleoid binding sites of the PomZ dimer are denoted as $\vec{x}_\text{clu}$ and $\vec{x}_\text{nuc}$, respectively. 

We expect that the position of a cluster, which is tethered to the nucleoid, does not change remarkably in the direction perpendicular to the nucleoid's surface because the cluster is unlikely to penetrate into the nucleoid's volume due to its large size and, on the other hand, cannot move far away from the nucleoid due to the tethering. 
Hence we neglect the forces that a PomZ dimer exerts on the cluster in the direction perpendicular to the nucleoid's surface by approximating the positions of the cluster and nucleoid binding site by their projections on the rectangular sheet representing the nucleoid, i.e.~$\vec{x}_\text{clu} = (x_\text{clu}, y_\text{clu})$, and $\vec{x}_\text{nuc} = (x_\text{nuc}, y_\text{nuc})$.

The rate for a PomZ dimer located at $\vec{x}_\text{nuc}$ to attach, with its second binding site, to any site of the cluster (``total attachment rate'') is then given by integration of $k_a(\vec{x}_\text{clu})$ over all possible cluster binding sites:
\begin{equation}
k_a^\text{tot} = \int_{\mathcal{A}_\text{clu}} k_{a}^0 \cdot \exp{\left[- \frac 1 2 \, \beta k (\vec{x}_\text{clu}-\vec{x}_\text{nuc})^{2}\right]} \text{d}\vec{x}_\text{clu} \approx 
\begin{cases}
	 k_a^0 \cdot \frac{2\pi}{\beta k}, & \text{if $\vec{x}_\text{nuc}\in \mathcal{A}_\text{clu}$}~,\\
	 0, & \text{otherwise}~.
\end{cases}
\end{equation}
Here, $\mathcal{A}_\text{clu}$ denotes the area on the nucleoid that is covered by the cluster. 
Since the Boltzmann factor decays quickly ($1/\sqrt{\beta k} = \SI{0.01}{\um}$ for the spring stiffness used, see Table~\ref{tab:sim_params}), we can neglect the boundaries of the region $\mathcal{A}_\text{clu}$ for $\vec{x}_\text{nuc}\in \mathcal{A}_\text{clu}$ and approximate the attachment rate of a PomZ dimer bound to the nucleoid outside of the cluster region by zero. 

Due to the fast decay of the exponential factor with an increase in $|\vec{x}_\text{clu}-\vec{x}_\text{nuc}|$, we can also save computation time by introducing a cut-off distance above which we set the attachment rate to zero.  
The cut-off is defined as the smallest distance $\Delta x = |\vec{x}_\text{clu}-\vec{x}_\text{nuc}|$ for which the attachment rate per lattice site, $k_a \cdot a^2$, is smaller than \SI{E-5}{\second^{-1}}. 
This value is chosen such that the average number of times this event occurs during the time the cluster takes to reach midcell ($\approx \SI{4800}{\second}$) is $\SI{0.048}{}$, i.e.~much lower than one. 

PomZ dimers are captured at the cluster until they are released into the cytosol upon ATP hydrolysis. 
In our model, we combine the different processes (ATP hydrolysis, conformational change in PomZ and detachment of PomZ-ADP into the cytosol) into one effective detachment process and denote the corresponding rate as the ATP hydrolysis rate $k_h$, which we assume to be independent of the degree of stretching of the dimer.

\subsection{Hopping of PomZ dimers on the nucleoid and the PomXY cluster}
In our model we assume that nucleoid-bound PomZ dimers can diffuse on the nucleoid and, when the dimer is attached to the cluster, also on the PomXY cluster with diffusion constants, $\Dnuc$ and $\Dclu$, respectively. 
These two diffusive processes are implemented as stochastic hopping events that occur with rate $k_\text{hop, nuc}^0 = \Dnuc /a^2$ and $k_\text{hop, clu}^0 = \Dclu /a^2$. 
More concretely, a nucleoid-bound PomZ dimer at site $(i,j)$, with $i$ denoting the lattice site in $x$-direction and $j$ in $y$-direction, can move to site $(i\pm 1,j)$ or $(i,j\pm 1)$ with hopping rate $k_{\text{hop, nuc}}^0$.
As we are using periodic boundary conditions in $y$-direction, a particle may also hop from site $(i, N_{\text{nuc},y})$ to the site $(i,1)$ and vice versa; here $N_{\text{nuc},y} = \wnuc/a$ denotes the number of sites along the $y$-axis.  
In $x$-direction we assume reflecting boundary conditions for the PomZ dimer movements. 

If a PomZ dimer is bound to both cluster and nucleoid, a hopping event leads to gain or loss in elastic energy.
In this case we multiply the constant hopping rates by exponential factors, which are chosen such that detailed balance holds (see Lansky et al.~[25]): 
\begin{linenomath}
\begin{align}
\label{Eq_Sup_RateDoubleHopping1}
k_\text{hop,clu}&=k_\text{hop,clu}^0\cdot\,\exp{\left[-\frac 1 4 \, \beta k \left[(\vec{x}_\text{clu, new}-\vec{x}_\text{nuc})^2-(\vec{x}_\text{clu, old}-\vec{x}_\text{nuc})^2\right]\right]}~,\\
\label{Eq_RateDoubleHopping2}
k_\text{hop,nuc}&=k_\text{hop,nuc}^0\cdot\,\exp{\left[-\frac 1 4 \, \beta k \left[(\vec{x}_\text{clu}-\vec{x}_\text{nuc, new})^2-(\vec{x}_\text{clu}-\vec{x}_\text{nuc, old})^2\right]\right]}~.
\end{align}
\end{linenomath}
The labels ``old'' and ``new'' refer to the positions of the binding sites before and after the hopping event. 

\subsection{Movement of the PomXY cluster} 
Cluster-bound PomZ dimers can exert forces, which lead to a net movement of the cluster.
Let us denote the position of the nucleoid binding site of the $i$-th PomZ dimer as $\vec{x}_{i,\text{nuc}}$, and the position of the cluster binding site as $\vec{x}_{i,\text{clu}} = \vec{x}_{c} + \Delta \vec{x}_{i,\text{clu}}$. 
With $\vec{x}_c = (x_c, y_c) \in \mathbb{R}^2$ we denote the position of the midpoint of the cluster.  
Here, we decomposed the position of a cluster binding site into two parts: the cluster position, $\vec{x}_c$, and an additional vector $\Delta \vec{x}_{i,\text{clu}}$. 
The reason for this decomposition is that we are interested in the equation of motion for the cluster position, $\vec{x}_c$, and, as long as the PomZ dimer does not diffuse on the nucleoid or the cluster, the two vectors $\vec{x}_{i,\text{nuc}}$ and $\Delta \vec{x}_{i,\text{clu}}$ are constant. 

A single PomZ dimer, bound to the nucleoid and cluster at fixed positions relative to both scaffolds, exerts a force $\vec{F}_i(t)$ on the cluster:
\begin{equation}
\vec{F}_i(t) = - k \left(\vec{x}_{i,\text{clu}}(t) - \vec{x}_{i,\text{nuc}}\right) = - k \left(\vec{x}_{c}(t) + \Delta \vec{x}_{i,\text{clu}} - \vec{x}_{i,\text{nuc}}\right)~.
\end{equation}
In a friction dominated regime, the sum over all forces exerted by $N_b$ cluster-bound PomZ dimers has to balance with the friction force acting on the cluster (friction coefficient $\gamma$):
\begin{equation}
\gamma \dot{\vec{x}}_{c}(t) = \sum_{i=1}^{N_b} \vec{F}_i(t) = -k \sum_{i=1}^{N_b}\left(\vec{x}_{c}(t) + \Delta\vec{x}_{i,\text{clu}} - \vec{x}_{i,\text{nuc}}\right)~.
\end{equation}
This equation is solved by separation of variables, yielding
\begin{equation}
\label{Eq_Sup_ClusterMovement}
\vec{x}_{c}(t) = (\vec{x}_{c}(t_0)-\vec{x}_f)\exp{\left(-\frac{N_b k}{\gamma}(t-t_0)\right)}+\vec{x}_f~,
\end{equation}
with 
\begin{equation}
\vec{x}_f = \frac{1}{N_b}\left(\sum_{i=1}^{N_b} \vec{x}_{i,\text{nuc}}-\Delta\vec{x}_{i,\text{clu}}\right)~.
\end{equation}
The initial time is denoted as $t_0$. 
We find that the cluster approaches the position $\vec{x}_f$, at which no net force is acting on the cluster, exponentially fast with characteristic time $t_\text{clu} = \gamma/(N_b k)$.

\subsection{\label{sec:SI_cyt_PomZ}Cytosolic PomZ distribution} 
To include the cytosolic PomZ dynamics in our model, we reduce the ATPase cycle to three processes: attachment of PomZ-ATP to the nucleoid, detachment of PomZ-ADP at the cluster, and nucleotide exchange.
We model the dynamics of PomZ-ATP and PomZ-ADP in the cytosol as one-dimensional reaction-diffusion equations as described in the following.
At the position of the cluster, $x_c(t)$, PomZ-ADP is released into the cytosol. 
Since the PomZ dynamics is a lot faster than the cluster dynamics, we can assume that the cluster is stationary, $x_c(t) = x_c$ on the time scale of the PomZ dynamics. 
The local increase in cytosolic PomZ-ADP at the cluster position due to detachment facilitated by ATP hydrolysis is approximated as a point source: $s_0 \delta(x-x_c)$. 
The constant $s_0$ depends on the hydrolysis rate $k_h$ and the amount of PomZ dimers bound to the cluster.
However, our final result, the normalized steady-state PomZ-ATP distribution will not depend on this constant. 
In the cytosol, PomZ-ADP exchanges ADP for ATP nucleotides with an effective rate $k_\text{ne}$. 
We assume that cytosolic PomZ in both nucleotide states diffuses with the same diffusion constant, $D_\text{cyt}$. 
However, only the ATP-bound form of PomZ can attach to the nucleoid with a rate $k_\text{on}$.
In total, we obtain the following coupled partial differential equations for the cytosolic PomZ-ATP ($c_T$) and PomZ-ADP ($c_D$) density:
\begin{linenomath}
\begin{subequations}
\begin{align}
\label{eq:SI_PomZ_ADP}
\partial_t c_D(x,t) &= D_\text{cyt} \partial_x^2 c_D(x,t) - k_\text{ne} c_D(x,t) + s_0 \delta(x-x_{c})\Theta(t)~, \\
\label{eq:SI_PomZ_ATP}
\partial_t c_T(x,t) &= D_\text{cyt} \partial_x^2 c_T(x,t) + k_\text{ne} c_D(x,t) - k_\text{on} c_T(x,t)~.
\end{align}
\end{subequations}
\end{linenomath}

We solved these two differential equations with no-flux boundary conditions for the stationary case. 
The solution for PomZ-ATP, given the cluster is at position $x_c$, reads:
\begin{linenomath}
\begin{alignat}{2}
c_T(x;x_c) &= \tilde{c}_1 \bigg[-\lambda_T \cosh{\left(\frac{L_1}{\lambda_T}\right)}\cosh{\left(\frac{L_2+x}{\lambda_T}\right)}\sinh{\left(\frac{L}{\lambda_D}\right)} + \nonumber \\ 
\label{eq:SI_PomZATP_sola}
&\hspace{10mm}+\lambda_D \cosh{\left(\frac{L_1}{\lambda_D}\right)}\cosh{\left(\frac{L_2+x}{\lambda_D}\right)}\sinh{\left(\frac{L}{\lambda_T}\right)}\bigg]~, && \hspace{2mm} \text{for} \hspace{3mm} -x_c\leq x\leq 0~,\\
c_T(x;x_c) &= \tilde{c}_1 \bigg[-\lambda_T \cosh{\left(\frac{L_2}{\lambda_T}\right)}\cosh{\left(\frac{L_1-x}{\lambda_T}\right)}\sinh{\left(\frac{L}{\lambda_D}\right)} + \nonumber \\ 
\label{eq:SI_PomZATP_solb}
&\hspace{10mm}+\lambda_D \cosh{\left(\frac{L_2}{\lambda_D}\right)}\cosh{\left(\frac{L_1-x}{\lambda_D}\right)}\sinh{\left(\frac{L}{\lambda_T}\right)}\bigg]~, && \hspace{2mm} \text{for} \hspace{3mm}0\leq x\leq l_\text{nuc} - x_c~,
\end{alignat} 
\end{linenomath}
with
\begin{equation}
\tilde{c}_1 = \frac{4 s_0 \lambda_T^2 e^{L(1/\lambda_D+1/\lambda_T)}}{D_\text{cyt} \left(\lambda_D^2 - \lambda_T^2\right)\left(e^{2L/\lambda_D}-1\right)\left(e^{2L/\lambda_T}-1\right)}~.
\end{equation}
We chose the coordinate system such that the cluster position, $x_c \in [0,l_\text{nuc}]$, is shifted to the origin. 
The lengths of the cluster-to-nucleoid end distances left and right of the cluster are given by $x_c$ and $l_\text{nuc} - x_c$, respectively. 
Furthermore, we defined the diffusive length scales for PomZ-ADP until it exchanges its ADP for ATP, and PomZ-ATP until it attaches to the nucleoid as $\lambda_D$ and $\lambda_T$, respectively:
\begin{equation}
\lambda_{D}=\sqrt{\frac{D_\text{cyt}}{k_\text{ne}}} \hspace{5mm} \text{and} \hspace{5mm} \lambda_{T}=\sqrt{\frac{D_\text{cyt}}{k_\text{on}}}~.
\end{equation}

The above solution for the cytosolic PomZ-ATP density holds true for $\lambda_T \neq \lambda_D$. 
If the two length scales are equal, $\tilde{c}_1$ becomes singular and hence this case needs to be considered separately. 
For $\lambda_D = \lambda_T \equiv \lambda$ the solution is given by:
\begin{linenomath}
\begin{alignat}{2}
\label{Eq_Sup_CytDiffSol2}
c_T(x; x_c) &= \tilde{c}_2\left[(2 L_1-x)\cosh{\left(\frac{2L_2+x}{\lambda}\right)}-x\cosh{\left(\frac{2L+x}{\lambda}\right)}+\right.\nonumber\\
&\hspace{7mm} \left. +~(2L+x)\cosh{\left(\frac{x}{\lambda}\right)}+(2L_2+x)\cosh{\left(\frac{2L_1-x}{\lambda}\right)}+\right.\nonumber\\
&\hspace{7mm} \left. +~4\lambda\cosh{\left(\frac{L_1}{\lambda}\right)} \cosh{\left(\frac{L_2+x}{\lambda}\right)}\sinh{\left(\frac{L}{\lambda}\right)}\right]~, && \text{for} \hspace{3mm} -x_c\leq x\leq 0~,\\
c_T(x;x_c) &= \tilde{c}_2\left[(2 L_2+x)\cosh{\left(\frac{2L_1-x}{\lambda}\right)}+x\cosh{\left(\frac{2L-x}{\lambda}\right)}+\right.\nonumber\\
&\hspace{7mm} \left.+~(2L-x)\cosh{\left(\frac{x}{\lambda}\right)}+(2L_1-x)\cosh{\left(\frac{2L_2+x}{\lambda}\right)}+\right.\nonumber\\
&\hspace{7mm} \left.+~4\lambda\cosh{\left(\frac{L_2}{\lambda}\right)} \cosh{\left(\frac{L_1-x}{\lambda}\right)}\sinh{\left(\frac{L}{\lambda}\right)}\right]~, && \text{for} \hspace{3mm} 0\leq x\leq l_\text{nuc}-x_c~,
\end{alignat}
\end{linenomath}
with
\begin{equation}
\tilde{c}_2 = \frac{s_0}{8D_\text{cyt} \sinh^2{\left(L/\lambda\right)}}~.
\end{equation}
With the analytical solution for the PomZ-ATP density in the cytosol, we can now define the attachment rate of a PomZ dimer to the nucleoid, accounting for the cytosolic PomZ distribution. 
We normalize the steady-state solution for $c_T(x; x_c)$ to obtain a probability density:
\begin{equation}
p_T(x;x_c)=\frac{\kon}{s_0} c_T(x;x_c)~.
\end{equation}
The rate for a PomZ dimer to attach to position $(x,y)$ on the nucleoid is then defined by:
\begin{equation}
\label{Eq_Sup_RateSingleAttachment2}
k_\text{on}(x,y;x_c) \equiv k_\text{on} \, p_T(x;x_c) p_T(y) = \frac{k_\text{on}}{w_\text{nuc}} \, p_T(x;x_c)~,
\end{equation}
with a uniform distribution, $p_T(y)$, along the short cell axis direction. 
Approximately, the attachment rate per lattice site is then given by this value multiplied with the lattice spacing $a$ squared. 

\begin{figure}[h!]
\centering
\includegraphics[width = 0.5\columnwidth]{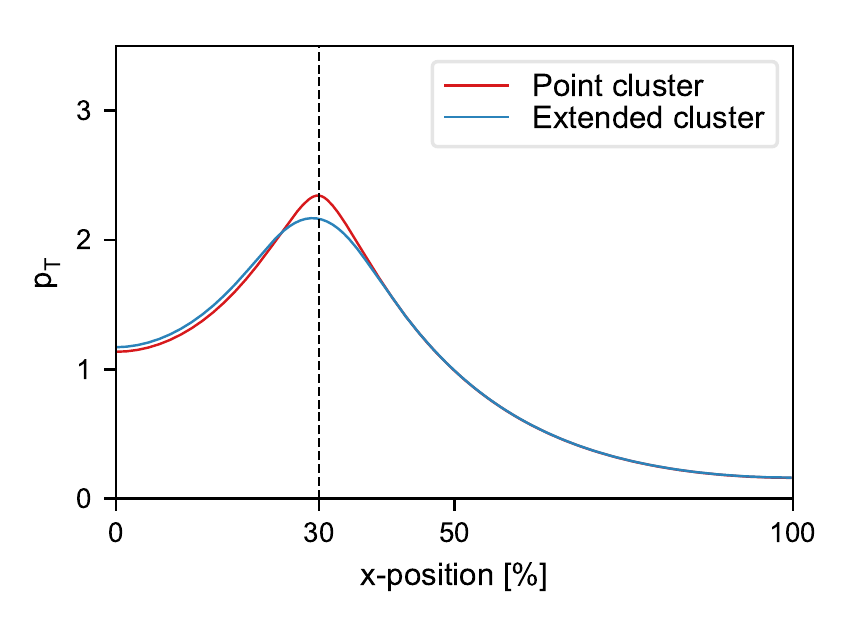}
\caption{\textbf{Cytosolic PomZ-ATP distribution.} Comparison of the steady-state solutions for the cytosolic PomZ-ATP density if release of PomZ-ADP into the cytosol at the PomXY cluster is modelled as a point source (red line) or a source with the same extension as the cluster (blue line).  
The cluster is at position $x_c = 30\%$ of nucleoid length.
We used the parameters shown in Table~\ref{tab:sim_params}. 
}
\label{Fig_Sup_CytDiff}
\end{figure}

In the presented derivation of the cytosolic PomZ-ATP density we reduced the cytosol to a one-dimensional line and the PomXY cluster to a point source. 
To investigate how the PomZ-ATP density changes when the cluster's extension is accounted for, we solved equations \ref{eq:SI_PomZ_ADP} and \ref{eq:SI_PomZ_ATP} with the Dirac delta distribution replaced by a Heaviside step function $ \Theta(x-x_c+\lclu/2)\Theta(x_c+\lclu/2-x)/\lclu$, numerically. 
We find that the steady-state solution for the PomZ-ATP density, when the cluster is included as a point source (Equations \ref{eq:SI_PomZATP_sola} and \ref{eq:SI_PomZATP_solb}), is a good approximation to the solution considering a one-dimensional cytosolic lane and an extended cluster (Figure~\ref{Fig_Sup_CytDiff}). 
The PomZ density profiles only deviate significantly in close proximity to the cluster, which can be attributed to the different shapes of the cluster used. 

\section{\label{sec:SI_params}Discussion of parameters used in the simulations}

\begin{table}[!ht]
\begin{tabular}{|l|c|c|c|} 
  \hline
  \textbf{Parameter} & \textbf{Variable} & \textbf{1D model} & \textbf{3D model}\\ 
  \hline
  Nucleoid length & $\lnuc$ & \SI{5.0}{\um} & \SI{5.0}{\um}\\
  Nucleoid width (circumference) & $\wnuc$ & - & \SI{2.2}{\um}\\
  \hline
  PomXY cluster length & $\lclu$ & \SI{0.7}{\um} & \SI{0.7}{\um}\\
  PomXY cluster width & $\wclu$ & - & \SI{0.7}{\um}\\
  \hline
    Effective spring stiffness of a PomZ dimer & $k$ & \SI{e4}{k_BT \um^{-2}}  & \SI{e4}{k_BT \um^{-2}}\\
    \hline
  Attachment rate of cytosolic PomZ to nucleoid & $k_\text{on}$ & \SI{0.1}{\second^{-1}} & \SI{0.1}{\second^{-1}}\\
  \hline 
  Attachment rate of nucleoid-bound PomZ to cluster (unstretched) & $k_a^0$ & \SI{500}{\second^{-1}\um^{-1}} & \SI{2.0e4}{\second^{-1}\um^{-2}}\\
  \hline
  Diffusion constant of PomZ on nucleoid & $\Dnuc$ & \SI{0.1}{\um^2\second^{-1}} & \SI{0.1}{\um^2\second^{-1}}\\
  Diffusion constant of PomZ on cluster  & $\Dclu$ & \SI{0.1}{\um^2\second^{-1}} & \SI{0.1}{\um^2\second^{-1}}\\
  \hline
  ATP hydrolysis rate of PomZ at the cluster & $k_h$ & \SI{1}{\second^{-1}} & \SI{1}{\second^{-1}}\\
  \hline
  Total number of PomZ dimers in the cell & $N$ & 100 & 100\\
  \hline
  Diffusion constant of the PomXY cluster in the cytosol & $\Dcluster$ & \SI{4e-4}{\um^2/s} & \SI{4e-4}{\um^2/s}\\
  \hline
  Diffusion constant of PomZ in the cytosol & $D_\text{cyt}$ & - & \SI{0.1}{\um^2 \second^{-1}},  \SI{0.5}{\um^2 \second^{-1}}\\
  \hline
  Nucleotide exchange rate of PomZ & $\kne$ & - & \SI{6}{\second^{-1}}\\
  \hline
  Lattice spacing & $a$ & \SI{0.01}{\um} & \SI{0.01}{\um}\\
  \hline
\end{tabular}
\caption{\textbf{Parameters used in the simulations.} 
The parameters used in the one-dimensional model are the same as in [10]. 
In the three-dimensional model we used the same parameter values when possible.
}
\label{tab:sim_params}
\end{table}
The total PomZ dimer number, the length and width of the cluster and the length of the nucleoid are chosen in accordance with experimental observations in \textit{M. xanthus} [4]. 
For the ATP hydrolysis rate we use a value of $\SI{1}{\per\second}$ as suggested by FRAP experiments where PomZ in the cluster is bleached [4]. 
The attachment rate of cytosolic PomZ to the nucleoid is approximated by literature values for the related Par system for chromosome and plasmid segregation (\SI{50}{\second^{-1}}~[13] and \SI{0.03}{\second^{-1}}~[12]). 
To get a double attachment rate that is comparable to the one-dimensional case, we chose $k_a^0$ such that the total attachment rate $k_a^\text{tot}$ is equal in the one-dimensional model studied previously [10] and the three-dimensional model studied here. 
This implies that the rate used in the one-dimensional model has to be multiplied by a factor of $\sqrt{{\beta k}/{2\pi}}$ to account for the additional dimension. 

The diffusion constant of PomZ on the nucleoid and on the PomXY cluster is approximated by the effective diffusion constant of ParA dimers on the nucleoid used in models for the Par system. 
However, these values vary a lot: \SI{0.001}{\um^2/s} -- \SI{1}{\um^2/s} [13, 26]. 
The friction coefficient of the cluster, $\gamma$, is related to its diffusion constant, $\Dcluster$, via Stokes-Einstein: $\gamma = k_BT/\Dcluster$.
We approximate the diffusion constant $\Dcluster$ by the corresponding literature values for plasmids. 
However, since the size of the PomXY cluster is larger than the typical size of a single plasmid, these values are only an upper bound for the diffusion constant of the Pom cluster: $\Dcluster \le \SI{e-3}{\um^2 \second^{-1}}$[13, 14]. 
The effective spring stiffness of the PomZ dimers, $k$, we approximate by the stiffness of a bond between a plasmid and the nucleoid via ParA dimers [26].

We approximate the nucleotide exchange rate of PomZ-ADP to PomZ-ATP by the corresponding rate for MinD proteins, $\SI{6}{\second^{-1}}$ [15].
Based on the large diffusion constant of Min proteins in the cytosol (on the order of \SI{10}{\um^2\second^{-1}}, [15]), and the fast dynamics of PomZ in the cytosol as observed in FRAP experiments [4], we expect the cytosolic diffusion constant of PomZ to be large. 
When we tested the effect of a non-homogeneous PomZ-ATP distribution in the cytosol on the cluster dynamics, we chose diffusion constants that are two orders of magnitudes smaller ($\SI{0.1}{\um^2/s}$ or $\SI{0.5}{\um^2/s}$). 

\section{\label{sec:SI_stochastic_simulation_details} Details on the stochastic simulation}

\subsection{Initial PomZ distribution}
Initially, all PomZ proteins are in the cytosol. 
Then we let the simulations run for a time $t_\text{min}$ of at least $10$ minutes with a fixed cluster position such that the PomZ proteins can approach their steady-state distribution. 
The time $t_\text{min}$ is chosen such that it is larger than the typical time scales for the PomZ dynamics, i.e.~the time scale for attachment to the nucleoid ($1/k_\text{on} = \SI{10}{s}$), the time scale for PomZ to explore the whole nucleoid by diffusion ($l_\text{nuc}^2/2D_\text{nuc} = \SI{125}{s}$), and the time scale for cluster-bound PomZ to detach ($1/k_h = \SI{1}{\s}$). 
After the initial time, $t_\text{min}$, recording starts. 
The cluster can now start to move or is kept at a fixed position (``stationary simulation'') during the entire simulation. 

\subsection{Gillespie algorithm}
We implemented our model using the Gillespie algorithm [10, 27], a stochastic simulation algorithm. 
Since the cluster position, $\vec{x}_c \in \mathbb{R}^2$, changes over time according to the forces cluster-bound PomZ dimers exert on the cluster (Equation \ref{Eq_Sup_ClusterMovement}), all rates that depend on the position of the cluster binding site of a PomZ dimer, depend on time. 
These include the attachment rate of a nucleoid-bound PomZ dimer to the cluster and the hopping rates of a PomZ dimer bound to the nucleoid and the cluster. 
However, if the cluster only moves slightly in one time step of the Gillespie algorithm, we can approximate the time-dependent rates as constant. 

To quantify the effect of the time dependence of the rates, let us consider a PomXY cluster with $N_b$ PomZ dimers bound to it such that a non-zero net force acts on the cluster. 
According to Equation \ref{Eq_Sup_ClusterMovement} the time scale for the cluster to relax to the force-free position, $\vec{x}_f$, is given by $t_\text{clu}= \gamma /(N_b k)$. 
The number of cluster-bound PomZ dimers, $N_b$, changes with the position of the cluster along the nucleoid.  
For a cluster positioned at $10\%$ of nucleoid length and one at mid-nucleoid the number of cluster bound PomZ dimers, as obtained from simulations, leads to $t_{\mathrm{clu}}\approx \SI{0.18}{\second}$ and $t_{\mathrm{clu}} \approx \SI{0.09}{\second}$, respectively (for the parameters as in Table \ref{tab:sim_params}).

Next, we consider the time step, $\Delta t$, until the next event happens in the Gillespie algorithm. 
The most frequent event is hopping of PomZ dimers on the nucleoid for the parameters we consider (Table~\ref{tab:sim_params}). 
Hence, the time step $\Delta t$ can be approximated by the time until a PomZ dimer bound to the nucleoid, hops on the nucleoid.  
The rate for the event that any of the nucleoid-bound PomZ dimers, $N_\text{nuc}$, hops in any of the four possible directions on the nucleoid (ignoring the boundaries) is given by $4 \, k^0_{\mathrm{hop,nuc}} N_{\text{nuc}}$. 
The typical time until the next event happens can then be approximated by the inverse of this rate. 
Again, the number of nucleoid-bound PomZ dimers, $N_\text{nuc}$, varies with the position of the cluster. 
For a cluster at $10\%$ of nucleoid length and one at mid-nucleoid, we get $\Delta t \approx \SI{3e-6}{\second}$ and $\Delta t \approx\SI{4e-6}{\second}$, respectively.
Since the typical time until the next event happens, $\Delta t$, is much smaller than the time scale for the movement of the cluster, $t_\text{clu}$, we can approximate all rates in the Gillespie algorithm as time-independent, which significantly improves the computational speed of the algorithm.

\section{Processing of simulated data}
\subsection{PomZ flux on the nucleoid}

\begin{figure*}[t!]
\centering
\includegraphics{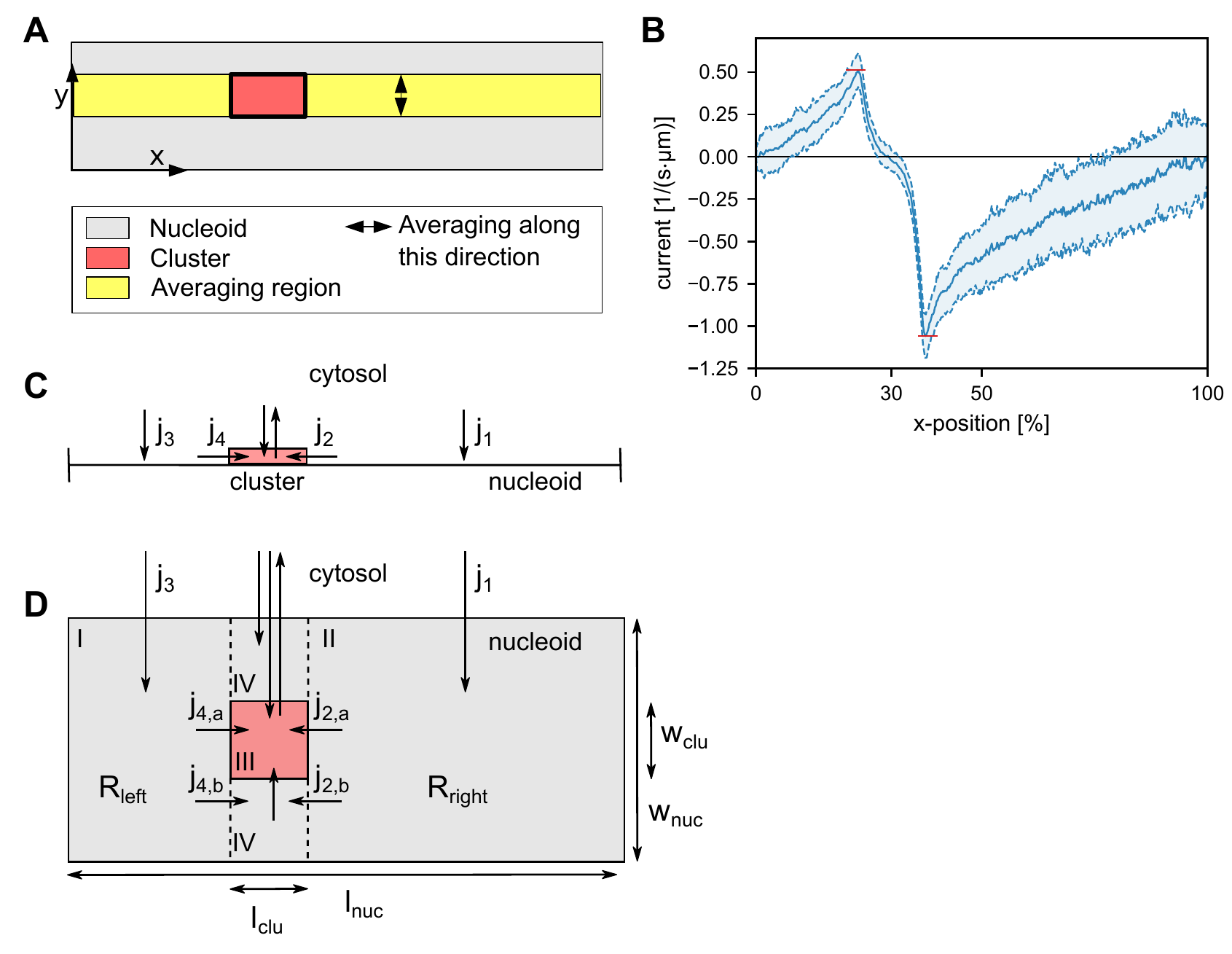}
\caption{\textbf{Flux of PomZ in the three-dimensional model geometry.} 
(A)~Sketch of the region used for averaging the flux. 
Only hopping events in the yellow region are taken into account when determining the PomZ flux profile along the long cell axis. 
The fluxes are averaged over the values in $y$-direction. 
(B)~Simulated flux profile of nucleoid-bound PomZ along the long cell axis with averaging performed as illustrated in (A). 
The flux into the cluster is given by the maximal value left, and the minimal value right of the cluster (red horizontal lines). 
We used the same parameters as in Table~\ref{tab:sim_params}.
(C)~In the one-dimensional model, nucleoid and cluster are incorporated as one-dimensional lattices. 
The cluster region is shown as a red rectangle. 
PomZ dimers attach to and diffuse on the nucleoid with reflecting boundary conditions at the nucleoid ends.
The arrows show the different PomZ fluxes from one region (cytosol, cluster and nucleoid left and right of the cluster) to another. 
(D)~Similar to (C), but the fluxes in the three-dimensional model geometry are shown. 
The grey region shows the nucleoid of size $l_\text{nuc} \times w_\text{nuc}$ and the red region the cluster of size $l_\text{clu} \times w_\text{clu}$. 
The area of the nucleoid regions left and right of the cluster are denoted by $R_\text{left}$ and $R_\text{right}$, respectively. 
}
\label{Fig_Sup_Measurement}
\end{figure*}

The PomZ flux along the nucleoid for a specific cluster position is determined by recording the PomZ flux at any time the cluster is in a small region ($\pm 0.5\%$ of nucleoid length) around the $x$-position of the cluster of interest.  
To obtain the flux of PomZ into the cluster along the long cell axis direction, the fluxes are averaged over the values in y-direction, but only considering the region of the nucleoid that corresponds to the extension of the cluster region along the long cell axis (yellow region in Figure~\ref{Fig_Sup_Measurement}A). 
Additionally, the data is averaged over an ensemble of about $100$ simulations. 
An example for such an averaged flux profile is shown in Figure \ref{Fig_Sup_Measurement}B. 
To obtain the difference in the PomZ fluxes into the cluster from each side along the long cell axis, the maximal / minimal values of the average flux profile left / right of the PomXY cluster are determined (red lines in Figure \ref{Fig_Sup_Measurement}B).
Finally, the two flux values of different signs are added together to obtain the flux difference.

\section{\label{sec:SI_flux_difference_3d}Flux difference into the cluster}
PomZ dimers detach from the nucleoid into the cytosol upon ATP hydrolysis, which breaks detailed balance and leads to a net flux of PomZ in the system. 
In the following we consider the fluxes of PomZ for the one- and three-dimensional model geometry (see Figure~\ref{Fig_Sup_Measurement}C,D).
Since the cluster is a lot less mobile than the PomZ dimers, we can approximate the cluster position as stationary. 
In this case, the fluxes in and out of each region (cytosol, cluster and nucleoid region left and right of the cluster) have to balance in the steady state. 
In the one-dimensional geometry, the fluxes of PomZ dimers to the nucleoid right and left of the cluster, $j_1$ and $j_3$, balance the fluxes into the cluster region, $j_2$ and $j_4$, respectively (Figure~\ref{Fig_Sup_Measurement}C). 
If PomZ is homogeneously distributed in the cytosol, the fluxes onto the nucleoid scale with the lengths of the respective nucleoid regions:
\begin{align}
j_1 &= k_\text{on} \frac{l_\text{nuc}-x_c-l_\text{clu}/2}{l_\text{nuc}} N_\text{cyt}~,\\
j_3 &= k_\text{on} \frac{x_c-l_\text{clu}/2}{l_\text{nuc}} N_\text{cyt}~, 
\end{align}
with $x_c$ the position of the cluster, and $N_\text{cyt}$ the number of PomZ dimers in the cytosol.
This results in the following formula for the flux difference of PomZ into the cluster \begin{equation}
j_\text{diff} = j_2-j_4 = j_1 - j_3 =  k_\text{on} N_\text{cyt} \left(1-\frac{2x_c}{l_\text{nuc}}\right).
\end{equation} 
The flux difference is proportional to the attachment rate of PomZ to the nucleoid, $k_\text{on}$, and the number of PomZ dimers in the cytosol, $N_\text{cyt}$. 
It is important to note, that $N_\text{cyt}$ also depends, among other parameters, on the position of the cluster $x_c$.

In the three-dimensional model geometry there are additional fluxes compared to the one-dimensional geometry if the cluster does not encompass the entire nucleoid circumference, i.e.~if $w_\text{clu} < w_\text{nuc}$ holds (see Figure~\ref{Fig_Sup_Measurement}D). 
Nucleoid-bound PomZ dimers in region I or II can leave these regions either by entering the cluster region and then attaching to the cluster, or by diffusing into the region in the extension of the cluster along the short cell axis (region IV in Figure~\ref{Fig_Sup_Measurement}D). 
In the latter case, the PomZ dimers can enter the cluster region along the short axis, diffuse back into the region they came from or diffuse past the cluster. 
In the steady state, the fluxes in and out of each region have to balance. 
For region I and II this implies:
\begin{align}
j_1 &= j_{2,a} + j_{2,b}~,\\
j_3 &= j_{4,a} + j_{4,b}~.
\end{align}
If we assume that the fluxes into the cluster region (region III) and into region IV scale with the extensions of the respective regions along the short cell axis, i.e.~$j_{2,a}/j_{2,b} = w_\text{clu}/(w_\text{nuc}-w_\text{clu})$ and similarly for $j_{4,a}$ and $j_{4,b}$, the flux difference into the cluster reads:
\begin{equation}
j_\text{diff} = j_{2,a} - j_{4,a} = k_\text{on} N_\text{cyt} \frac{w_\text{clu}}{w_\text{nuc}}\left(1-\frac{2x_c}{l_\text{nuc}}\right)~,
\end{equation}
which agrees with the formula for the one-dimensional system if the cluster is ring-shaped, i.e.~$w_\text{clu} = w_\text{nuc}$.
This analytical expression fits well with our simulation results for a ring-shaped cluster (see Figure~3D).
However, if the cluster does not cover the full nucleoid's width, it deviates from the simulation results. 
This deviation can be attributed to the fact that PomZ dimers that diffuse into region IV are not absorbed here, but can diffuse back into regions I or II. 
Hence, the fluxes into the cluster are larger than the values obtained from the simple estimate we used before. 
For the fluxes into the cluster from the right, we have
\begin{equation}
\frac{j_{2,a}}{j_{2,b}} > \frac{w_\text{clu}}{w_\text{nuc}-w_\text{clu}}~.
\end{equation} 
Furthermore, PomZ dimers can pass the cluster by diffusion from region I to II or vice versa. 
However, this flux only matters if the cluster is small both in length and width. 

\begin{figure*}[h!]
\centering
\includegraphics{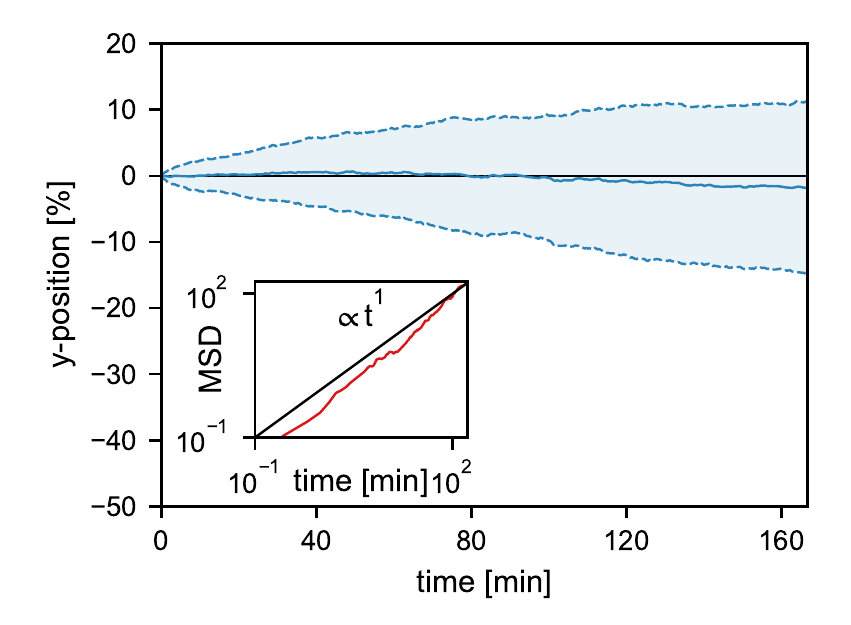}
\caption{\textbf{Cluster movement along short cell axis direction.} 
Average cluster trajectory in $y$-direction using an ensemble of $100$ simulations (solid blue line). 
The shaded region indicates one standard deviation above and below the average trajectory. 
The inset shows the mean-square displacement, which increases linearly in time, indicating diffusive motion.}
\label{Fig_Sup_Movement_y}
\end{figure*}

\section{Cluster movement along short cell axis direction}

In addition to the cluster dynamics along the long cell axis, as discussed in the main text, we also considered the dynamics along the short cell axis. 
Because of the rotational symmetry the clusters do not have a preferred direction along this axis, on average. 
However, both diffusive and persistent, unidirectional motion is conceivable. 
We expect persistent movement if PomZ diffusion on the nucleoid is slow compared to the cluster dynamics, such that there is a delay between the cluster movement and the PomZ gradient reaching its steady state [10, 14, 28 - 30].
In this case, the initial direction of the cluster's movement along the nucleoid's circumference is chosen stochastically by the interactions of the PomZ dimers with the cluster. 
Once the cluster started to move in one direction, it is more likely to continue in this direction, because the asymmetry in the PomZ density gradient is maintained.
In contrast, for fast PomZ diffusion on the nucleoid, we expect an approximately symmetric PomZ distribution around the cluster, resulting in equal likelihood for the cluster to move in each direction, suggesting diffusive motion. 
For the parameter set we considered (Table S1) the clusters show diffusive behavior in $y$-direction as indicated by a mean-square displacement that grows linearly in time (Figure~\ref{Fig_Sup_Movement_y}, inset).

\hspace{5cm}

\section{Dynamics of two Pom clusters}
\begin{figure}[h!]
\centering
\includegraphics{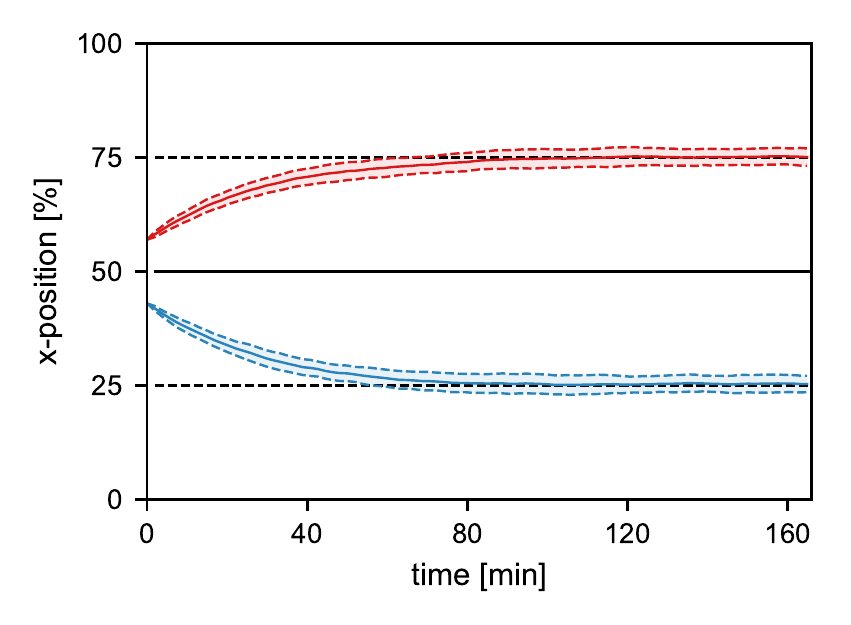}
\caption{\textbf{Two clusters are localized at the one- and three-quarter positions.} 
Averaged positions of the two clusters along the $x$-direction (solid lines, ensemble of $100$ runs). 
The shaded regions indicate the average cluster positions plus and minus one standard deviation. 
Initially, the two clusters are positioned side by side along the long cell axis such that the left edge of one and the right edge of the other cluster are positioned at mid-nucleoid. 
We used the parameter set given in Table S1.}
\label{fig:two_clusters}
\end{figure}

Motivated by equidistant positioning of plasmids by ParAB\textit{S} systems we investigated the dynamics of two Pom clusters in the realistic three-di\-men\-sio\-nal cell geometry. 
We find that two clusters localize at the one- and three-quarter positions along the nucleoid, i.e.~at equidistant positions (Figure~\ref{fig:two_clusters}A).

Flux-based positioning of two clusters can be heuristically explained as follows (see also [13, 14, 21]): 
If two protein clusters approach each other, the PomZ density and fluxes on the nucleoid between the clusters are reduced. 
Since the clusters move into the direction of the highest PomZ flux, two clusters effectively repel each other, which separates the clusters from each other.
Moreover, the average positions of the two clusters and the average PomZ density on the nucleoid has to be symmetric with respect to the mid-nucleoid plane.
Hence there is no net flux of nucleoid-bound PomZ at mid-nucleoid, on average. 
Therefore, in the stationary state, we can map the system to one that consists of two subsystems with half the size of the original system and each subsystem contains one cluster. 
The positioning mechanism previously discussed for one cluster explains midcell localization in each of the subsystems, which corresponds to the one- and three-quarter positions on the nucleoid.